\documentclass[]{cit_lab_mfr}
\usepackage{hyperref}
\usepackage{cleveref}
\usepackage{verbatim}

\usepackage{wrapfig}  
\usepackage{graphicx}
\usepackage{floatrow}
\usepackage{subcaption}
\usepackage{listings}
\usepackage{algorithm}
\usepackage{wrapfig}
\usepackage{subfig}
\usepackage{epigraph}
\usepackage{subcaption} %

\usepackage[toc,page,header]{appendix}

\usepackage{minitoc}

\newcommand{\modelhome}[1]{\href{#1}{\textsuperscript{$\dagger$}}}
\newcommand{\ie}{\textit{i.e.}}
\newcommand{\eg}{\textit{e.g.}}

\title{Douyin Multimodal Embedding Model Technical Report}

\affiliation{%
\parbox{\textwidth}{\centering
\textbf{ByteDance Douyin Search Multimodal Team}\quad \textbf{Renmin University of China GSAI}
}}

\renewcommand{\thefootnote}{\fnsymbol{footnote}}
\renewcommand{\thefootnote}{\arabic{footnote}}  

\usepackage{xspace}

\abstract{

Multimodal representation learning is a cornerstone of modern AI. By encoding multimodal queries and targets into vectors, it powers industrial applications such as search and recommendation, and increasingly underpins modern agents. Real-world platforms with complex modalities and massive-scale content, such as Douyin, Xiaohongshu, and YouTube, demand two capabilities simultaneously, efficiency under billion-scale indexing and fine-grained semantic discrimination for hard matching. Existing MLLM embedding models often struggle to jointly satisfy both requirements. Contrastive models are efficient but rely on pair-level supervision that is too coarse for fine-grained distinctions, while CoT-based models improve discrimination at the cost of explicit generation that is impractical to serve online. We present Douyin Multimodal Embedding (DME), a model that combines both strengths to meet the industrial demand for efficient and fine-grained representation. Specifically, DME is trained in two stages. Stage 1 performs large-scale contrastive pre-training that establishes a unified multimodal embedding space with broad modality and task coverage. Stage 2 supplements semantic sufficiency, the property that an embedding is grounded in retrieval-relevant evidence and preserves fine-grained counterpart-side semantics, through two complementary mechanisms. Evidence-Grounded Typed Latent Reasoning organizes retrieval evidence via hidden-space latent reasoning, and Cross-Conditional Reconstruction enforces counterpart-side semantics via cross-directional autoregressive reconstruction. Cross-Conditional Reconstruction is used only during training, and the latent tokens remain inside a single encoder forward pass and introduce only marginal query-side overhead. The generative supervision further makes DME embeddings information-complete, in that the input content can be recovered from them, which we quantify as an interpretable measure of semantic sufficiency and use to guide optimization in industrial settings. On MMEB-v2, DME achieves state-of-the-art results at comparable scales for both the 2B and 9B variants (\textbf{74.8} and \textbf{78.4}), with particularly strong performance on video and visual-document retrieval. In production, DME delivers a 2.92\% relative improvement in overall score on Douyin's in-house offline evaluation set, and has been deployed across a range of real-world Douyin scenarios such as generative search, image search, and AI search. Online A/B testing on Douyin search further confirms a 0.1\% Lifetime (LT) gain.
}
\date{\today}
\begin{document}
\maketitle

\begingroup 
\renewcommand{\thefootnote}{\fnsymbol{footnote}}
\endgroup

\begin{figure*}[ht]
    \centering
    \includegraphics[width=1\linewidth]{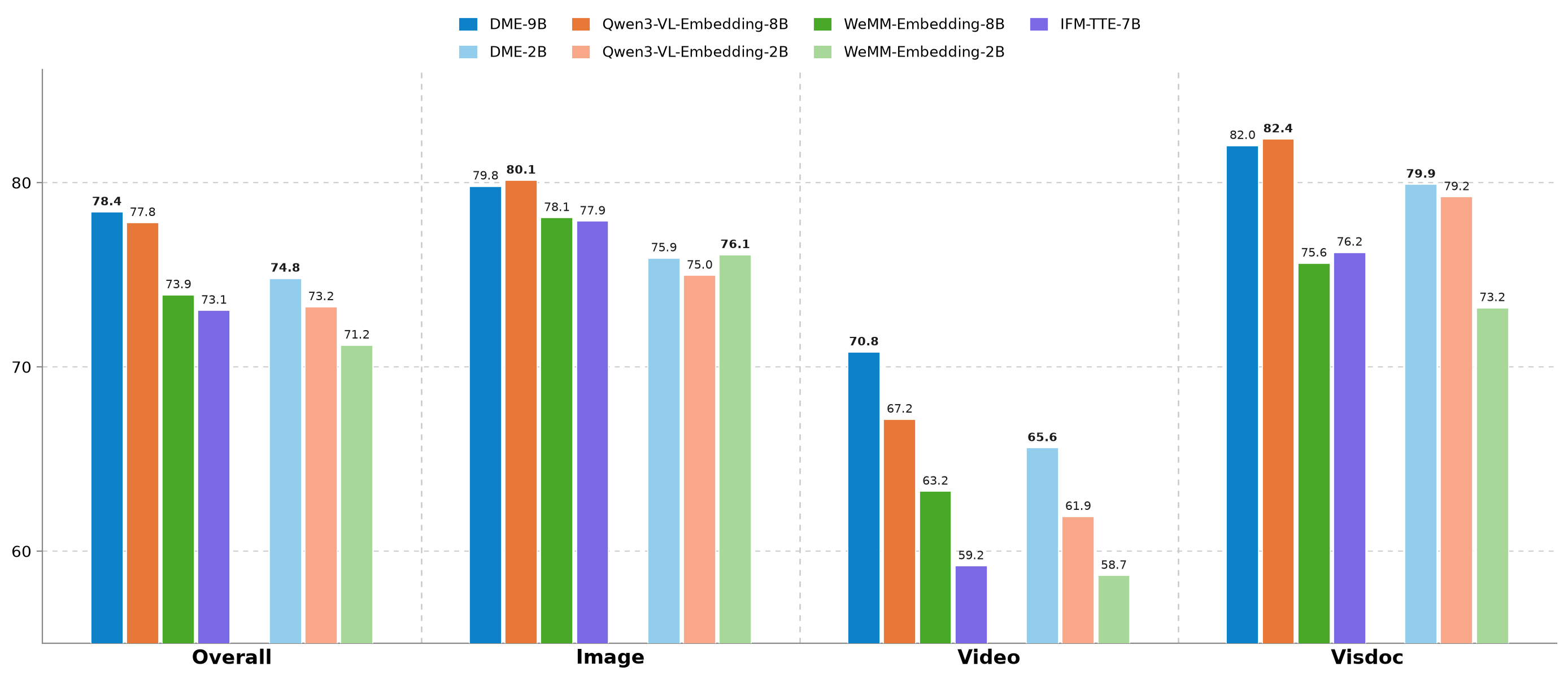}
    \caption{\textbf{Performance Comparison.} Overall and per-domain (Image, Video, VisDoc) results on MMEB-v2. At both the 2B and 9B scales, DME consistently outperforms other embedding models with especially large gains on video and visual-document retrieval.}
  \label{fig:performance}
\end{figure*}

\section{Introduction}
\label{sec:introduction}

Multimodal retrieval has become a foundational infrastructure of the AI era. Beyond search and recommendation, it is now also relied upon to retrieve external knowledge for retrieval-augmented generation and to serve as the tool through which agents perceive the outside world. These new callers also raise the bar: unlike short keyword queries, AI search and agents issue multi-constraint instructions and expect evidence that can support subsequent reasoning, not merely plausibly relevant candidates. In these systems, both user intents and candidate contents may appear as text, images, videos, visual documents, or their arbitrary mixtures. This requirement is further amplified on large-scale video and image--text platforms such as Douyin, Xiaohongshu, and YouTube, where real-world content is massive in scale and highly heterogeneous. User queries may be expressed through natural language, images, videos, or mixed-modality inputs, while candidate items may contain visual frames, video dynamics, OCR text, captions, metadata, and other textual signals. A practical retrieval model must therefore satisfy two requirements simultaneously: it should provide broad modality and task coverage under massive-scale indexing, while still preserving fine-grained semantic discrimination for difficult query--document matching.

Early vision-language contrastive models such as CLIP and ALIGN demonstrate that large-scale paired supervision can induce strong cross-modal alignment~\cite{CLIP,ALIGN}. More recently, frontier MLLMs such as Qwen3-VL and Qwen3.5 have become increasingly attractive as foundation backbones for multimodal embedding, as they provide stronger visual-language understanding, instruction following, long-context modeling, and multimodal reasoning ability~\cite{Qwen3VL,Qwen35Blog}. Building on these foundation models, recent MLLM-based embedding systems adapt generative MLLMs into unified multimodal retrievers over text, images, videos, visual documents, and mixed-modality inputs~\cite{E5-V,GME,mme5,u-marvel,VLM2VecV2,Qwen3VLEmbedding}, and have become the mainstream paradigm for universal multimodal retrieval.

The dominant recipe for training these MLLM-based retrievers is still contrastive learning. Positive query--document pairs are pulled together, while negatives are pushed apart in the embedding space~\cite{infonce,CLIP,E5,GTE}. This paradigm is scalable, compatible with offline corpus encoding, and well aligned with industrial vector-search systems. However, we argue that contrastive MLLM embedders face an inherent tension: they inherit the strong perception and reasoning potential of generative MLLMs, yet use the model mainly as an encoder optimized by pair-level similarity supervision. Such supervision tells the model which instances should be close, but never why. It neither identifies the local evidence that grounds a relevance decision, nor requires the final embedding to preserve the fine-grained counterpart-side semantics that distinguish a truly relevant pair.

This limitation becomes more critical in large-scale industrial multimodal retrieval. Many real queries are not simple category or caption matching requests; they may involve multiple constraints over objects, attributes, actions, OCR text, temporal order, or spatial relations. At the same time, hard negatives often share the same global scene or topic with the positive document, and differ only in a small visual region, a key frame, a local text span, or a subtle semantic condition~\cite{unime}. Recent benchmarks have therefore moved beyond conventional image--text retrieval toward video, visual-document, moment, and reasoning-intensive retrieval~\cite{VLM2VecV2,ColPali,visrag,MMBright}, where such fine-grained, locally-grounded matching is especially demanding. In these scenarios, an embedding should not merely encode global similarity; it should be formed from retrieval-relevant evidence and preserve enough semantic detail to distinguish partially matched candidates.

Recent studies have explored how to reintroduce reasoning into multimodal retrieval representations. Some methods generate embedding-centric reasoning traces or let the model adaptively decide whether to emit a reasoning chain before the embedding token~\cite{ThinkThenEmbed,TRACE}, while others use retrieval-oriented reinforcement learning to drive reasoning-augmented generative embeddings~\cite{UMER1,EmbedRL}. These methods show that reasoning can improve fine-grained retrieval, but many rely on explicit textual reasoning
or reranking-style computation. Such designs are powerful but sacrifice efficiency, making them difficult to deploy as the main retrieval encoder in billion-scale online systems.

\begin{figure*}[tp]
    \centering
    \includegraphics[width=1\linewidth]{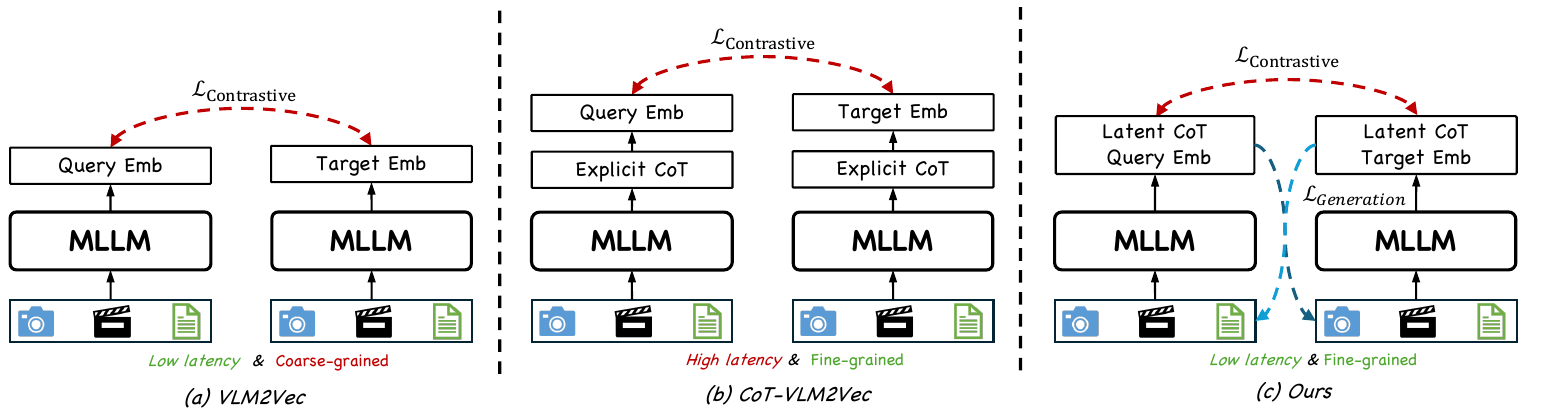}
\caption{\textbf{Comparison with prior works.} (a) Contrastive MLLM embedders (e.g., VLM2Vec) encode each side in a single pass, yielding \emph{low latency} but only \emph{coarse-grained} representations. (b) CoT-based embedders prepend explicit reasoning before the embedding, improving \emph{fine-grained} discrimination at the cost of \emph{high latency}. (c) DME performs latent reasoning together with cross-conditional generative supervision, achieving \emph{fine-grained} representations while retaining the \emph{low latency} of a bi-encoder.}
  \label{fig:comparison}
\end{figure*}

This creates two practical design requirements for industrial multimodal retrieval. First, the model should acquire evidence-aware reasoning ability without turning online retrieval into an explicit generation or reranking process, so that adding reasoning capacity introduces only marginal query-encoding overhead. 
Second, the representation itself must retain enough fine-grained counterpart-side detail to separate hard negatives, since even the right evidence can collapse into a single vector that drops such detail. Crucially, both abilities should be obtained without explicit generation or additional model passes, so that retrieval preserves the standard dense-vector interface.

Motivated by this industrial gap, we introduce Douyin Multimodal Embedding (DME), a two-stage multimodal embedding model designed to meet the two requirements above by combining the efficiency of contrastive MLLM embedders with the fine-grained semantic modeling ability of reasoning-augmented retrievers. Stage 1 performs large-scale contrastive pre-training over heterogeneous multimodal data, establishing a broad unified embedding space across text, images, videos, visual documents, and mixed-modality inputs. Stage 2 then improves the semantic sufficiency of the learned embeddings. We use semantic sufficiency to describe a stronger requirement for retrieval representations: an embedding should be not only pairwise aligned with relevant instances, but also grounded in retrieval-relevant evidence and capable of preserving fine-grained counterpart semantics from the matched query or document.

To achieve this goal, Stage 2 introduces two complementary mechanisms. The first is Evidence-Grounded Typed Latent Reasoning, which injects a controllable hidden-space reasoning process into the embedding model. Instead of generating explicit long-form CoT, DME uses latent tokens to organize retrieval computation before the final embedding readout. Anchor tokens first localize retrieval-relevant evidence from multimodal inputs, such as text spans, image regions, OCR fragments, or video keyframes. Typed latent reasoning states then organize the localized evidence into
retrieval-specific roles, including semantic localization, positive alignment,
and negative rejection. Teacher-generated trajectories may additionally contain
summarization states. Finally, a readout representation fuses the latent reasoning states and evidence representations into the final retrieval embedding. This design introduces the structure of evidence localization, latent reasoning, and embedding readout into multimodal retrieval, while preserving the efficiency of a bi-encoder model.

The second Stage-2 mechanism is Cross-Conditional Reconstruction, which combines Next Token Prediction (NTP) and Multi-Token Prediction (MTP). Reconstruction and generative supervision have long been used to improve representations, from masked language and image modeling to contrastive-captioning objectives~\cite{bert,MAE,CoCa}. Recent multimodal embedding studies further show that content reconstruction or joint generative-retrieval training can encourage MLLMs to compress richer semantic information into embedding tokens~\cite{CoCoA,CREM}. Inspired by this direction, DME uses the retrieval embedding itself as a semantic bottleneck for generation-oriented supervision. Given a positive query--document pair, the query embedding is used as a prefix condition to reconstruct document-side textual content, and the document embedding is symmetrically used to reconstruct query-side textual content. MTP further extends this signal by predicting multiple future tokens rather than only the next token, encouraging the embedding to capture longer-range semantic information~\cite{MultiTokenPrediction,DeepSeekV3}. These objectives provide fine-grained token-level supervision during training, without requiring additional decoding during retrieval inference. A further benefit of this generative supervision is that the token-level content of the input can be recovered from its embedding. We quantify this as a representation completeness measure, giving an interpretable measure of semantic sufficiency and an explicit signal to guide representation optimization.

Together, the two Stage-2 components address the two requirements above. Evidence-Grounded Typed Latent Reasoning improves how an embedding is formed: it encourages the model to look at the right evidence, organize retrieval-specific latent states, and fuse them into the final vector. Cross-Conditional Reconstruction improves what an embedding must preserve: it requires the vector representation to retain enough counterpart-side semantics to support conditional textual reconstruction. Both mechanisms are used to improve the embedding model itself. At inference time, DME remains an efficient bi-encoder retriever: each query or document is independently encoded into a dense vector, and retrieval is performed by vector similarity without explicit reasoning generation, agentic tool use, or cross-encoder scoring.

The main contributions of this report are summarized as follows:
\begin{itemize}
    \item We present Douyin Multimodal Embedding (DME), a two-stage multimodal embedding model for large-scale industrial multimodal retrieval. DME combines large-scale contrastive pre-training with semantic sufficiency learning, aiming to preserve bi-encoder efficiency while improving fine-grained multimodal understanding.

    \item We introduce two complementary Stage-2 mechanisms for semantic sufficiency: Evidence-Grounded Typed Latent Reasoning, which grounds embeddings in localized evidence through anchor-based hidden-space reasoning, and Cross-Conditional Reconstruction, which uses NTP and MTP to inject counterpart-side semantic supervision, so that the input content can be recovered from its embedding.

    \item We train and evaluate DME at both 2B and 9B scales. On MMEB-v2, DME-2B and DME-9B achieve state-of-the-art results among models of comparable sizes, with especially strong performance on video and visual-document retrieval (Figure~\ref{fig:performance}), while the latent reasoning tokens introduce only minor query-encoding latency overhead. We further show, through a token-level recovery analysis that we formalize as a representation completeness measure, that DME embeddings are information-complete, giving direct evidence of semantic sufficiency. Deployed in Douyin's real-world retrieval system, DME delivers a 2.92\% relative improvement in overall score on the in-house offline evaluation set, with consistent gains across all cross-modal retrieval directions, and has been adopted across a range of scenarios such as generative search, image search, and AI search. Online A/B testing on Douyin search further confirms a 0.1\% Lifetime (LT) gain.
\end{itemize}

The report is organized as follows. Section~\ref{sec:dme-overview} provides an overview of the DME architecture, retrieval setting, embedding extraction, and two-stage learning pipeline. Section~\ref{sec:multi-stage-learning-framework} describes the multi-stage learning framework in detail, including large-scale contrastive pre-training and the two Stage-2 semantic sufficiency mechanisms. Section~\ref{sec:experiments} presents the experimental setup, main results, and analysis. Section~\ref{sec:conclusion} concludes the report and discusses future directions.
\section{Related Work}
\label{sec:relatedwork}

\subsection{Multimodal Embedding Models}

Vision-language contrastive models such as CLIP and ALIGN first showed that large-scale paired supervision can induce strong cross-modal alignment~\cite{CLIP,BLIP,blip2,SigLIP,SigLIP2,ALIGN}. Building on stronger MLLM backbones~\cite{Qwen25-VL,Qwen3VL,Qwen35Blog}, a growing line of work adapts generative MLLMs into unified multimodal embedders, including E5-V, GME, mmE5, VLM2Vec-V2, U-MARVEL, PDF-VLM2Vec and Qwen3-VL-Embedding~\cite{E5-V,GME,mme5,u-marvel,VLM2VecV2, pdfvlm2vec,Qwen3VLEmbedding}. These models extend retrieval from image--text pairs to videos, visual documents, and mixed-modality inputs, and are typically trained with contrastive learning over positive and negative query--document pairs~\cite{infonce,E5,GTE}, which is efficient, offline-encodable, and well aligned with industrial vector search. 
A parallel line of industrial systems scales this paradigm to production search
and recommendation, such as Pailitao-VL and
SAIL-Embedding~\cite{PailitaoVL,SAILEmbedding}, and the MOON series, which
progressively targets large-scale training~\cite{MOON}, modality
balance~\cite{MOON2}, and reasoning ability~\cite{MOON3}. These efforts confirm
the practical value of multimodal embedding under billion-scale traffic, yet
they remain centered on contrastive pair-level supervision.
However, such pair-level supervision only specifies which instances should be close, without modeling why they are relevant or which fine-grained evidence an embedding should preserve. DME follows the same efficient bi-encoder paradigm but supplements this pair-level objective with evidence-grounded and generative supervision to improve semantic sufficiency.

\subsection{Reasoning-Augmented Retrieval}
Chain-of-thought prompting and reinforcement learning have substantially
strengthened reasoning in multimodal large language models, through explicit
intermediate rationales, structured step-by-step solving, and reasoning-oriented
reward optimization~\cite{MMCoT,LLaVaCoT,VisionR1,R1Onevision,VCSTaR}. This
progress motivates a natural question for retrieval: whether such reasoning
ability can be transferred into the representations used for matching, rather
than kept only in free-form text generation.
A recent line of work reintroduces reasoning into retrieval representations. Think-Then-Embed generates embedding-centric reasoning traces before extracting a representation, and TRACE adaptively decides whether to generate a reasoning chain before emitting an embedding token~\cite{ThinkThenEmbed,TRACE}. UME-R1 explores reasoning-driven generative embeddings with retrieval-oriented reinforcement learning~\cite{UMER1}, while Embed-RL uses a frozen embedder as a reward model to train a reasoner that produces retrieval-useful chains of thought~\cite{EmbedRL}. These methods confirm that reasoning improves fine-grained retrieval, but most of them rely on explicit textual reasoning, additional generation, agentic tool use, or reranking-style computation, which increase query-encoding latency and complicate corpus indexing in billion-scale online systems. 
In contrast, DME performs reasoning entirely in the hidden space through a small number of retrieval-specific latent tokens, keeping inference close to a standard dense retriever while still acquiring evidence-aware reasoning ability.

\subsection{Generative and Reconstruction-based Representation Learning}
Reconstruction and generative objectives have long been used to strengthen representations, from masked language and image modeling to contrastive-captioning objectives~\cite{bert,MAE,CoCa}. In the multimodal embedding setting, recent studies show that content reconstruction or joint generative-retrieval training can push MLLMs to compress richer semantic information into embedding tokens~\cite{CoCoA,CREM}, and multi-token prediction has been shown to encourage representations that capture longer-range semantics~\cite{MultiTokenPrediction,DeepSeekV3,deepseekv4}. DME builds on this direction but uses the retrieval embedding itself as a semantic bottleneck for cross-conditional decoding, reconstructing counterpart-side content from the query and document embeddings. This not only injects fine-grained token-level supervision during training, but also makes the embedding decodable back to its input, which we use as a quantitative measure of semantic sufficiency rather than as an inference-time generation step.

\section{DME Overview}
\label{sec:dme-overview}

\subsection{Task Formulation and Retrieval Setting}
\label{sec:task-formulation}

DME is designed for instruction-aware universal multimodal retrieval. We use the term ``document'' to denote a generic retrievable item, which may be a text passage, an image, a video or a mixed-modality object. Similarly, a query may also be text-only, vision-only, video-based, or composed of multiple modalities. This setting covers a broad family of retrieval tasks, including text-to-text retrieval, text-to-image retrieval, image-to-text retrieval, video retrieval, visual document retrieval, multimodal question answering, classification-as-retrieval, and moment-level retrieval~\cite{MMEB,VLM2VecV2,Qwen3VLEmbedding,ColPali,visrag}.

Formally, let $\mathcal{X}$ denote the space of multimodal instances. Each instance $x \in \mathcal{X}$ is represented as a token sequence constructed from one or more modalities:
\begin{equation}
    x = [x^{\mathrm{text}}, x^{\mathrm{image}}, x^{\mathrm{video}}],
\end{equation}
where absent modalities are omitted. A retrieval task is specified by a natural-language instruction $\iota$, which defines the relevance criterion between a query and a document. For example, the instruction may ask the model to retrieve an image matching a textual description, find a document that answers a question, classify an input by retrieving the correct label, or locate a video segment corresponding to a temporal event.

We organize training and evaluation data as a collection of task-specific retrieval datasets:
\begin{equation}
    \mathcal{D} = \{\mathcal{D}_m\}_{m=1}^{M},
    \qquad
    \mathcal{D}_m = (\iota_m, \mathcal{Q}_m, \mathcal{C}_m, \mathcal{R}_m),
\end{equation}
where $\iota_m$ is the task instruction, $\mathcal{Q}_m$ is the query set, $\mathcal{C}_m$ is the candidate corpus, and $\mathcal{R}_m$ contains relevance annotations. For each query $q_i \in \mathcal{Q}_m$, the relevance annotation is written as
\begin{equation}
    \mathcal{R}_m(q_i) = (\mathcal{P}_i, \mathcal{N}_i),
\end{equation}
where $\mathcal{P}_i \subset \mathcal{C}_m$ denotes relevant documents and $\mathcal{N}_i \subset \mathcal{C}_m$ denotes irrelevant or hard-negative documents. The learning objective is to produce a scoring function that ranks documents in $\mathcal{P}_i$ above those in $\mathcal{N}_i$:
\begin{equation}
    s_\theta(q_i, d^+ \mid \iota_m) > s_\theta(q_i, d^- \mid \iota_m),
    \quad
    d^+ \in \mathcal{P}_i,\ d^- \in \mathcal{N}_i.
\end{equation}

DME follows a bi-encoder retrieval setting. Given an instruction $\iota$, a query $q$, and a document $d$, the model independently encodes the query and document into dense vectors:
\begin{equation}
    \mathbf{z}_q
    =
    E_{\theta}^{q}
    \left(
        T_q(\iota,q)
    \right),
    \qquad
    \mathbf{z}_d
    =
    E_{\theta}^{d}
    \left(
        T_d(\iota,d)
    \right),
\label{eq:bi-encoder-extraction}
\end{equation}
where $T_q$ and $T_d$ are query-side and document-side input templates, and $E_{\theta}^{q}$ and $E_{\theta}^{d}$ denote the corresponding embedding extraction procedures. Both extractors share the same MLLM backbone and embedding projection, while allowing side-specific retrieval-token layouts and readout functions. The resulting embeddings are $\ell_2$-normalized.

The relevance score is computed by inner product, which is equivalent to cosine similarity after normalization:
\begin{equation}
    s_\theta(q,d\mid\iota)
    =
    \mathbf{z}_q^\top \mathbf{z}_d.
\label{eq:retrieval-score}
\end{equation}

At inference time, all document embeddings in the corpus are computed offline and stored in an approximate nearest-neighbor index. Given an online query, DME performs a single query-side encoding pass and retrieves candidates by vector similarity:
\begin{equation}
    \operatorname{Retrieve}(q, \mathcal{C})
    =
    \operatorname{TopK}_{d \in \mathcal{C}}
    \ s_\theta(q, d \mid \iota).
\end{equation}
This retrieval setting is critical for large-scale deployment because it avoids pairwise cross-attention between the query and every candidate document. It also separates training-time supervision from inference-time computation.


\subsection{Model Architecture and Embedding Extraction}
\label{sec:model-architecture-embedding}

DME is built on a generative multimodal large language model as a shared embedding backbone. The backbone receives an instruction-aware serialized sequence consisting of text tokens, visual tokens, and retrieval-specific tokens. The visual tokens may come from images, video frames, visual documents, or other visual inputs. Following recent MLLM-based embedding models~\cite{E5-V,GME,MMEB,VLM2VecV2,Qwen3VLEmbedding}, DME converts the generative backbone into a dense retriever by extracting compact vector representations from retrieval-side hidden states.

Given a retrieval instruction $\iota$ and a multimodal instance $x$, DME constructs the input sequence as:
\begin{equation}
    \mathbf{u}
    =
    T(\iota, x)
    =
    [
    \mathbf{u}_{\mathrm{inst}};
    \mathbf{u}_{x};
    \mathbf{u}_{\mathrm{ret}}
    ],
\end{equation}
where $\mathbf{u}_{\mathrm{inst}}$ denotes the instruction tokens, $\mathbf{u}_{x}$ denotes the serialized multimodal content, and $\mathbf{u}_{\mathrm{ret}}$ denotes retrieval-specific tokens appended after the input content. Since these retrieval tokens are placed at the end of the sequence, their hidden states can attend to the preceding multimodal context under the causal attention mechanism of the MLLM backbone.

Let $F_{\theta}$ denote the MLLM backbone. Given the serialized sequence $\mathbf{u}$, the model produces final-layer hidden states:
\begin{equation}
    \mathbf{H}
    =
    F_{\theta}(\mathbf{u})
    =
    [
    \mathbf{h}_1,
    \mathbf{h}_2,
    \ldots,
    \mathbf{h}_{|\mathbf{u}|}
    ].
\end{equation}
DME extracts a readout representation $\mathbf{r}$ from the retrieval-specific hidden states and projects it into the embedding space:
\begin{equation}
    \mathbf{z}
    =
    \operatorname{norm}
    \left(
        W_{\mathrm{emb}}
        \mathbf{r}
    \right),
\end{equation}
where $W_{\mathrm{emb}}$ is the embedding projection matrix and $\operatorname{norm}(\cdot)$ denotes $\ell_2$ normalization. The concrete construction of $\mathbf{r}$ differs across training stages and is summarized in Section~\ref{sec:two-stage-learning-pipeline}.

The side-specific extraction functions $E_{\theta}^{q}$ and $E_{\theta}^{d}$ are defined in Eq.~\eqref{eq:bi-encoder-extraction}. They share the same MLLM backbone and embedding projection, but may use different retrieval-token layouts and readout functions. In Stage 2, both query and document inputs may use modality-aware anchor tokens for evidence grounding, while typed latent reasoning is introduced on the query side. The resulting normalized embeddings are compared using the score in Eq.~\eqref{eq:retrieval-score}.

\subsection{Two-Stage Learning Pipeline}
\label{sec:two-stage-learning-pipeline}


\begin{figure*}[tp]
    \centering
    \includegraphics[width=1\linewidth]{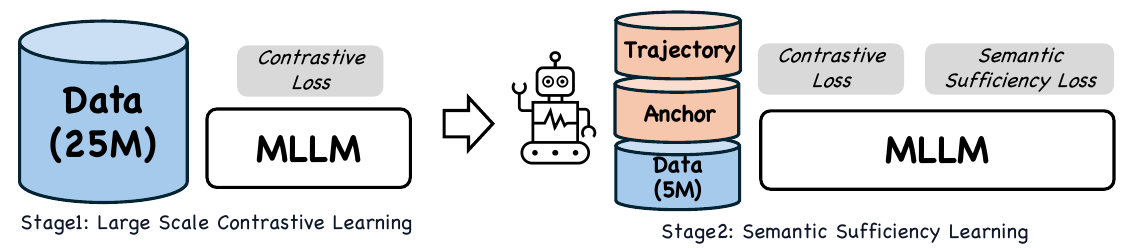}
    \caption{\textbf{Overview of the DME two-stage training pipeline.}
    In Stage 1, the MLLM backbone is
    trained on 25M multimodal query-target pairs with a contrastive loss to
    establish a scalable bi-encoder embedding space.
    In Stage 2, the model is further
    refined on a smaller, higher-quality set of 5M examples augmented with
    teacher-generated CoT supervision. Stage 2 is optimized
    jointly by the contrastive loss and a semantic sufficiency loss, the latter
    combining Evidence-Grounded Typed Latent Reasoning and Cross-Conditional
    Reconstruction. Cross-Conditional Reconstruction is used only during training. The latent tokens remain inside a single-pass encoder and introduce only marginal query-side overhead, while retrieval is still performed through standard dense-vector similarity.}
    \label{fig:pipeline}
  \label{fig:pipeline}
\end{figure*}

DME is trained in two major stages, as shown in Figure~\ref{fig:pipeline}. Stage 1 builds a broad multimodal embedding space through large-scale contrastive pre-training, while Stage 2 improves the semantic sufficiency of the learned representations through evidence-grounded latent reasoning and cross-conditional reconstruction. The detailed objectives are introduced in Section~\ref{sec:multi-stage-learning-framework}; here we summarize the overall pipeline and how each stage changes the retrieval readout.

\textbf{Stage 1: large-scale contrastive pre-training.}
In the first stage, DME uses a standard embedding token as the retrieval-specific token:
\begin{equation}
    \mathbf{u}_{\mathrm{ret}}^{(1)}
    =
    [
    \texttt{<emb>}
    ].
\end{equation}
The readout representation is obtained from the final hidden state of this token:
\begin{equation}
    \mathbf{r}^{(1)}
    =
    \mathbf{h}_{\texttt{<emb>}},
    \qquad
    \mathbf{z}^{(1)}
    =
    \operatorname{norm}
    \left(
        W_{\mathrm{emb}}
        \mathbf{r}^{(1)}
    \right).
\end{equation}
This simple readout allows DME to scale contrastive pre-training over heterogeneous query--document pairs, including text, image, video, visual document, and mixed-modality data. The goal of this stage is to establish broad modality alignment and a stable initial retrieval geometry.

\begin{figure*}[t]
    \centering
    \includegraphics[width=1\linewidth]{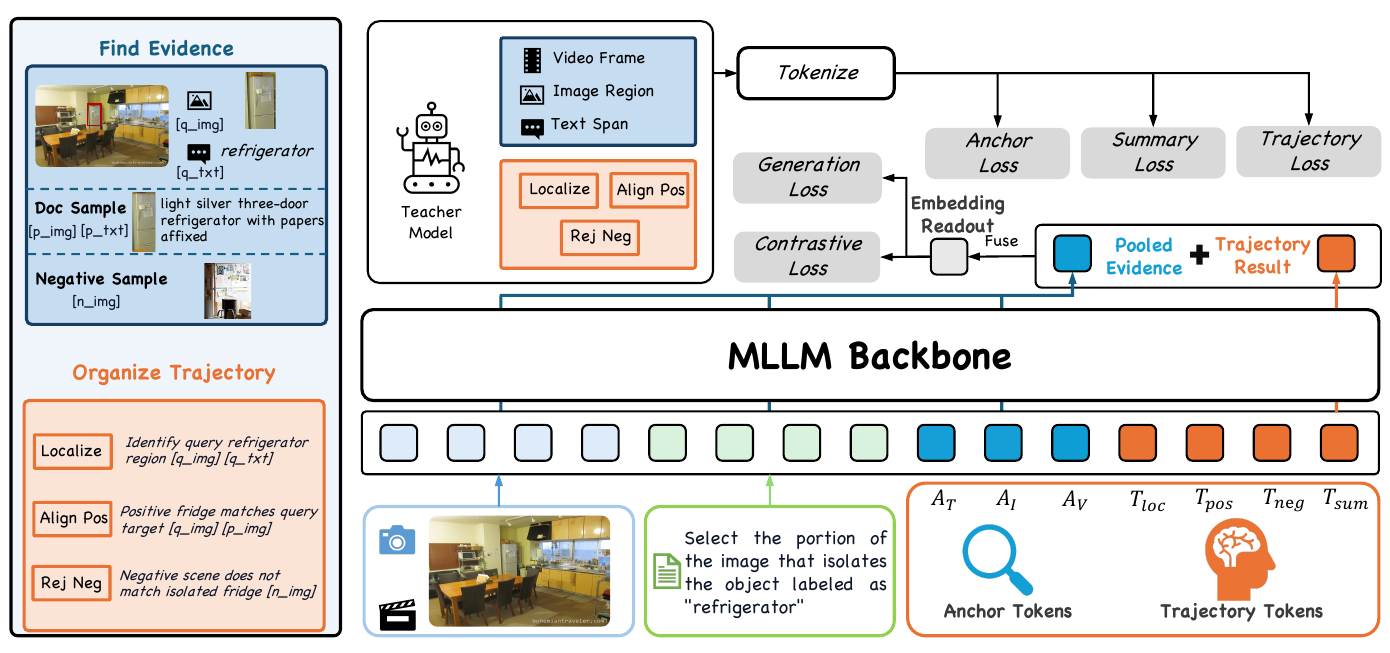}
\caption{\textbf{Stage 2: Semantic Sufficiency Learning.}
A teacher model decomposes a query--positive--negative triplet into localized multimodal evidence and typed retrieval states. These signals supervise modality-aware anchor tokens and query-side latent states, whose terminal state is combined with pooled evidence to form the retrieval representation. Cross-Conditional Reconstruction provides an additional NTP/MTP objective during training, shown as the generation loss in the figure. At evaluation time, DME performs a single encoder forward pass without explicit textual CoT generation or reconstruction.}
\label{fig:stage2}
\end{figure*}

\textbf{Stage 2: semantic sufficiency learning.}
Stage 2 retains the same normalized embedding interface while introducing two complementary mechanisms (Figure~\ref{fig:stage2}). Both query and document inputs may use modality-aware anchor tokens for evidence grounding. The query side additionally uses a small number of typed latent states, whose terminal state contributes to the query readout, while the document side retains a standard document readout. Stage 2-A uses these representations for Evidence-Grounded Typed Latent Reasoning, and Stage 2-B further uses the readout representations as conditioning prefixes for NTP/MTP-based reconstruction during training. The detailed formulations are provided in Sections~\ref{sec:stage2a-typed-latent-reasoning} and~\ref{sec:stage2b-cross-conditional-self-decoding}.

\section{Multi-Stage Learning Framework}
\label{sec:multi-stage-learning-framework}

DME follows a two-stage training design. The first stage performs large-scale contrastive pre-training to build a general multimodal embedding space across heterogeneous retrieval tasks. The second stage refines this embedding space with two complementary mechanisms:
Evidence-Grounded Typed Latent Reasoning, which grounds the retrieval representation
in selected multimodal evidence and typed latent states, and Cross-Conditional
Reconstruction, which applies NTP/MTP to enforce fine-grained counterpart semantics
in the readout representation. We describe the training objective of each stage and then summarize the final joint objective.

\subsection{Training Data and Supervision Sources}
\label{sec:training-data-supervision}

DME uses different supervision sources across training stages. Data scaling is central to establishing the model's fundamental multimodal representation capability, while later stages progressively increase supervision quality and task specificity. Stage 1 mainly relies on large-scale weakly supervised contrastive data to learn broad cross-modal semantic alignment. Stage 2 shifts toward higher-quality, instruction-formatted supervised data to improve retrieval performance across specialized and complex multimodal scenarios. Stage 2-A further introduces synthetic structured supervision for evidence grounding and typed latent reasoning. We describe these data sources separately in this section, and leave the detailed objectives to the following subsections.

\paragraph{Stage-1 contrastive data.}

The primary goal of Stage 1 is to establish fundamental multimodal embedding capabilities through data scaling. We mainly rely on large-scale weakly supervised contrastive data across text, image, and video to learn basic semantic alignment across modalities.

Beyond public data, we further introduce in-house proprietary and synthetic data to complement existing corpora in terms of data quality, scenario coverage, and temporal understanding. High-quality image captions help improve fine-grained image--text alignment, while long-video captions and video--text alignment data enhance the model's ability to model temporal dynamics and complex visual scenes. With dense frame-level and clip-level supervision, these video data further strengthen temporal and visual representation learning. In total, Stage 1 is trained on approximately 25M query--document pairs.

\paragraph{Stage-2 contrastive data.}

The objective of Stage 2 is to comprehensively improve retrieval capability across specialized scenarios and complex multimodal tasks. To this end, we construct a diverse mixture of instruction-formatted retrieval data, where each example explicitly specifies the retrieval intent and task context. This design encourages the model to better understand different retrieval objectives, rather than relying only on implicit query--document relevance signals.

The dataset includes the training set from MMEB-v2~\cite{VLM2VecV2}, supplemented by a wide array of public multimodal retrieval and question-answering datasets. These datasets cover image--text retrieval, video--text retrieval, visual-document retrieval, question answering, classification-style retrieval, and mixed-modality retrieval. All examples are normalized into a unified $(\text{instruction}, \text{query}, \text{positive})$ format. Sampling ratios are balanced across task families to prevent large datasets or individual task types from dominating the training distribution, which improves generalization and stability across different retrieval scenarios.

\paragraph{Stage-2A structured CoT supervision data.}
The supervision signals for Stage 2-A are generated from the original query--positive--negative retrieval samples using Seed-2.0-Pro~\cite{seed2} as the teacher model. For each raw sample, we first normalize it into a structured triplet. The teacher then generates item-level structured anchor records for the query and positive document. When a hard negative is available, we also generate a negative-side structured anchor record, which is mainly used to construct rejection-oriented trajectory supervision rather than as an anchor-grounding target.

We use the term \emph{structured anchor record} to distinguish the supervision data from the model anchor tokens introduced later. A structured anchor record is not a learnable token. Instead, it is a teacher-generated item-level annotation that contains modality-specific evidence entries and a short \emph{local summary}. The evidence entries may be text spans, image regions, or video frame references. The local summary is generated jointly with the evidence entries; if it is missing or invalid, we derive a fallback summary from the first valid text, image, or video evidence entry. For text-only and classification-style samples, a lightweight heuristic path can directly use short text or label semantics as the local summary.

The teacher outputs structured annotations rather than model tokens. Each record contains query-side and positive-side evidence annotations, an optional negative-side evidence record used to construct rejection supervision, item-level local summaries, and a short typed retrieval trajectory. Structured evidence may refer to text spans, image regions, or video frame indices. During preprocessing, these fields are aligned to the serialized Qwen3.5 input to obtain token-, patch-, or frame-level supervision targets. The local summaries and trajectory-state descriptions are encoded offline into cached semantic targets.

The typed trajectory describes the relation among the query, positive document, and optional negative document. Each step contains a retrieval role, references to previously generated evidence entries, and a short state description. The supported roles include \texttt{localize}, \texttt{align\_pos}, \texttt{reject\_neg}, and \texttt{summarize}. Query and positive evidence provide anchor-grounding supervision, whereas negative-side evidence is used only to construct rejection-oriented trajectory states.

\begin{tcolorbox}[
    colback=blue!5,
    colframe=blue!65!black,
    title={Structured Supervision for Stage 2-A},
    fonttitle=\bfseries,
    boxrule=0.8pt,
    arc=2pt
]
\small
\begin{tabular}{p{0.24\linewidth}p{0.68\linewidth}}
\texttt{query\_anchor} &
Query-side text, image, or video evidence entries with structured locations and semantic labels. \\

\texttt{pos\_anchor} &
Positive-side evidence used together with query evidence for anchor-grounding supervision. \\

\texttt{neg\_anchor} &
Optional negative-side evidence used to construct \texttt{reject\_neg} states; not used as an anchor hit target. \\

\texttt{local\_summary} &
A short item-level semantic summary encoded offline for summary-conditioned supervision. \\

\texttt{traj} &
A short typed trajectory whose steps contain a retrieval role, referenced evidence ids, and a state description. \\
\end{tabular}
\end{tcolorbox}

\paragraph{Data filtering.}

To ensure stable and consistent training signals, we systematically filter the training data by removing empty samples, corrupted media, low-quality captions, low-resolution images, and near-duplicate examples. This process reduces noisy supervision and improves the consistency of cross-modal alignment.

\paragraph{Negative construction and in-batch sampling.}

Negative construction plays an important role in contrastive training. In Stage 1, we mainly rely on in-batch negatives and increase the effective negative space by using large batch sizes, so that each query is contrasted against a larger set of candidate distractors at every optimization step.

In Stage 2, in addition to in-batch negatives, we incorporate hard negatives when available. These include annotated distractors, retrieval-mined candidates, and semantically similar examples from the same task family. For retrieval-mined hard negatives, we first use existing models to perform offline retrieval and scoring over large candidate pools, and then select samples that are semantically close to the query but do not match the ground-truth relevance. Compared with random negatives, these hard negatives better resemble real retrieval distractors and encourage the model to distinguish between superficially related and truly relevant content.

To mitigate false-negative noise in large-scale contrastive learning, we filter potential pseudo negatives by measuring their similarity to the positive document. Negatives whose similarity scores are overly high are removed from the candidate pool, reducing the risk of optimizing against samples that may also be valid answers. During training, we also adopt task-balanced in-batch mixed sampling: samples from the same data source are drawn consecutively according to a controlled ratio, while different task types are mixed within a batch. This strategy preserves the stability of same-source training and introduces cross-task and cross-modal contrastive signals, leading to more robust unified representations.

\subsection{Stage 1: Large-Scale Contrastive Pre-training}
\label{sec:stage1-contrastive-pretraining}

Stage 1 aims to build a broad and stable multimodal embedding space before applying the Stage-2 semantic sufficiency objectives. At this stage, DME is trained as a standard bi-encoder retriever with large-scale heterogeneous query--document pairs. The model uses the Stage-1 readout token $\texttt{<emb>}$ introduced in Section~\ref{sec:two-stage-learning-pipeline}, and does not yet use anchor tokens, typed latent states, or reconstruction supervision. This design makes Stage 1 simple and scalable, while providing a strong initialization for the more structured Stage-2 training.

The contrastive data mixture used in this stage is summarized in Section~\ref{sec:training-data-supervision}.
Given an instruction-aware query template $T_q$ and a document template $T_d$, DME encodes each pair using the side-specific extraction functions defined in Eq.~\eqref{eq:bi-encoder-extraction}:
\begin{equation}
    \mathbf{z}_{q_i}
    =
    E_{\theta}^{q}
    \left(
        T_q(\iota_i,q_i)
    \right),
    \qquad
    \mathbf{z}_{d_i^+}
    =
    E_{\theta}^{d}
    \left(
        T_d(\iota_i,d_i^+)
    \right).
\label{eq:stage1-pair-encoding}
\end{equation}
The embeddings are $\ell_2$-normalized and compared using the score in Eq.~\eqref{eq:retrieval-score}.

\paragraph{Contrastive objective.}
For each query $q_i$, let $\mathcal{D}_i^-$ denote its negative document set. In practice, $\mathcal{D}_i^-$ can be constructed from other documents in the same mini-batch and from the data mixture. The Stage-1 contrastive objective encourages the query representation to be closer to the positive document than to negative documents:
\begin{equation}
    \mathcal{L}_{\mathrm{S1}}^{\mathrm{CL}}
    =
    -
    \frac{1}{B}
    \sum_{i=1}^{B}
    \log
    \frac{
        \exp
        \left(
            s_{\theta}(q_i, d_i^+ \mid \iota_i) / \tau_{\mathrm{CL}}
        \right)
    }{
        \exp
        \left(
            s_{\theta}(q_i, d_i^+ \mid \iota_i) / \tau_{\mathrm{CL}}
        \right)
        +
        \sum_{d^- \in \mathcal{D}_i^-}
        \exp
        \left(
            s_{\theta}(q_i, d^- \mid \iota_i) / \tau_{\mathrm{CL}}
        \right)
    },
\label{eq:stage1-cl}
\end{equation}
where $B$ is the mini-batch size, $\tau_{\mathrm{CL}}$ is the temperature parameter, and $s_{\theta}(\cdot,\cdot)$ is the vector-similarity score defined in Section~\ref{sec:task-formulation}.

\paragraph{Role of Stage 1.}
Stage 1 provides the foundation for the full DME training pipeline. It aligns heterogeneous modalities into a common embedding space, gives the model broad task coverage, and produces a stable retrieval geometry for large-scale nearest-neighbor search. However, the supervision in this stage remains pair-level: it optimizes whether a query and a document should be close, but does not explicitly model which evidence supports the match or whether the embedding preserves fine-grained counterpart semantics. These limitations motivate the two Stage-2 components introduced in the following subsections.

\subsection{Stage 2-A: Evidence-Grounded Typed Latent Reasoning}
\label{sec:stage2a-typed-latent-reasoning}

Stage 2-A introduces an evidence-grounded latent reasoning module that encourages the retrieval representation to depend on evidence supporting the query--document match. Stage-1 contrastive learning supervises the final similarity score, but does not explicitly determine where the encoder should attend or what retrieval role an intermediate hidden state should represent. Stage 2-A addresses this limitation through modality-aware anchor tokens and a small number of typed query-side latent states.

The structured CoT records described in Section~\ref{sec:training-data-supervision} provide three forms of supervision: localized evidence targets, item-level local summaries, and typed retrieval-state descriptions. The design follows a compact loop from evidence grounding to retrieval optimization: anchors identify where to look, summary targets constrain what the selected evidence means, typed latent states organize retrieval-specific information, and the resulting representation is used by the contrastive objective.

\paragraph{Evidence grounding.}
For readability, we suppress the sample index in the main text. Let $s\in\{q,d\}$ denote the query or document side, and let $m(r)$ denote the modality associated with anchor $r$. For anchor hidden state $\mathbf{a}_{s,r}$ and content-token hidden state $\mathbf{x}_{s,j}$ from the same modality, DME computes:
\begin{equation}
    p_{s,r}(j)
    =
    \operatorname{softmax}_{
        j\in\mathcal{I}_{m(r)}(s)
    }
    \left(
        \frac{
            \left(
                W_q^{a}\mathbf{a}_{s,r}
            \right)^{\top}
            \left(
                W_k^{a}\mathbf{x}_{s,j}
            \right)
        }{
            \sqrt{d_a}
        }
    \right),
\end{equation}
where $\mathcal{I}_{m(r)}(s)$ denotes the content tokens belonging to the modality of anchor $r$. The corresponding side-level evidence pool is:
\begin{equation}
    \mathbf{e}_{s,\mathrm{pool}}
    =
    \operatorname{Mean}_{r}
    \left(
        \sum_{
            j\in\mathcal{I}_{m(r)}(s)
        }
        p_{s,r}(j)
        \mathbf{x}_{s,j}
    \right).
\label{eq:stage2a-evidence-pool}
\end{equation}

Teacher-generated evidence targets supervise the anchor distributions through an evidence hit loss $\mathcal{L}_{\mathrm{hit}}$. Cached local-summary embeddings additionally supervise the semantic content of the pooled evidence through $\mathcal{L}_{\mathrm{sum}}$. The evidence assignment, assignment balancing, and summary-conditioned objectives are detailed in Appendix~\ref{app:stage2a-details}.

\paragraph{Typed latent supervision.}
The query side additionally contains a small number of latent states associated with retrieval roles such as semantic localization, positive alignment, and negative rejection. Semantic states are aligned with cached teacher-state embeddings, positive-alignment states are trained to retrieve the corresponding positive document, and rejection states are trained to prefer the positive document over a paired hard negative. We summarize these objectives as:
\begin{equation}
    \mathcal{L}_{\mathrm{typed}}
    =
    \lambda_{\mathrm{sem}}
    \mathcal{L}_{\mathrm{sem}}
    +
    \lambda_{\mathrm{align}}
    \mathcal{L}_{\mathrm{align}}
    +
    \lambda_{\mathrm{reject}}
    \mathcal{L}_{\mathrm{reject}}.
\label{eq:stage2a-typed-loss}
\end{equation}
The definitions of the three terms are provided in Appendix~\ref{app:stage2a-details}.

\paragraph{Evidence-enhanced readout.}
Let $\mathbf{h}_{q,R}^{\mathrm{traj}}$ denote the terminal query-side latent state. The query readout combines this terminal state with the query-side evidence pool. The document side uses its standard readout state together with the document-side evidence pool:
\begin{equation}
    \mathbf{r}_q
    =
    \mathbf{h}_{q,R}^{\mathrm{traj}}
    +
    \alpha\,
    \operatorname{sg}
    \left(
        W_e
        \mathbf{e}_{q,\mathrm{pool}}
    \right),
    \qquad
    \mathbf{r}_d
    =
    \mathbf{h}_{d,\mathrm{read}}
    +
    \alpha\,
    \operatorname{sg}
    \left(
        W_e
        \mathbf{e}_{d,\mathrm{pool}}
    \right),
\label{eq:stage2a-readout}
\end{equation}
where $\operatorname{sg}(\cdot)$ denotes stop-gradient. The final query and document embeddings are:
\begin{equation}
    \mathbf{z}_s
    =
    \operatorname{norm}
    \left(
        W_{\mathrm{emb}}
        \mathbf{r}_s
    \right),
    \qquad
    s\in\{q,d\}.
\label{eq:stage2a-embedding}
\end{equation}

The Stage-2A objective is:
\begin{equation}
    \mathcal{L}_{\mathrm{S2A}}
    =
    \lambda_{\mathrm{hit}}
    \mathcal{L}_{\mathrm{hit}}
    +
    \lambda_{\mathrm{sum}}
    \mathcal{L}_{\mathrm{sum}}
    +
    \mathcal{L}_{\mathrm{typed}}.
\label{eq:stage2a-objective}
\end{equation}

This formulation links evidence selection to retrieval without introducing explicit textual reasoning. The latent states remain inside the encoder forward pass, while the structured teacher targets provide supervision for evidence grounding and retrieval-oriented latent representations.

\subsection{Stage 2-B: Cross-Conditional Reconstruction with NTP and MTP}
\label{sec:stage2b-cross-conditional-self-decoding}

While Stage 2-A improves \emph{how} an embedding is formed, Stage 2-B constrains \emph{how much} counterpart-side information the final vector must retain. Under pair-level similarity supervision alone, the fine-grained generative understanding inherited from the MLLM backbone tends to degrade, and the embedding is never required to preserve the counterpart semantics that distinguish a relevant pair. Stage 2-B closes this gap by turning the retrieval embedding itself into a semantic bottleneck that must be sufficient to \emph{reconstruct} the counterpart-side textual content during training. Concretely, the query embedding is used as a prefix condition to decode the document-side text, and the document embedding is symmetrically used to decode the query-side text. This cross-conditional reconstruction injects fine-grained, token-level supervision into the embedding, while leaving the inference-time retrieval interface unchanged. We instantiate it with two complementary objectives: Next Token Prediction (NTP), which supervises the next token, and Multi-Token Prediction (MTP), which additionally supervises multiple future tokens so that the embedding captures longer-range counterpart semantics.

\paragraph{Embedding as a prefix condition.}
Stage 2-B reuses the Stage-2 readout representation introduced in Section~\ref{sec:stage2a-typed-latent-reasoning}. For a query $q$ and a document $d$ under instruction $\iota$, the readout vectors $r_q$ and $r_d$ produce the retrieval embeddings $z_q=\mathrm{norm}(W_{\text{emb}}r_q)$ and $z_d=\mathrm{norm}(W_{\text{emb}}r_d)$. We use the \emph{pre-normalization} embedding directly as the decoding condition, without any additional projection:
\begin{equation}
\tilde{\mathbf{z}}_q = W_{\text{emb}}\,r_q,
\qquad
\tilde{\mathbf{z}}_d = W_{\text{emb}}\,r_d .
\label{eq:s2b-prefix}
\end{equation}
The vector $\tilde{\mathbf{z}}_q$ (resp. $\tilde{\mathbf{z}}_d$) is then placed at the first position of the decoding sequence as a single prefix token, so that all subsequently decoded tokens attend to it under the causal attention of the shared backbone $F_\theta$. Because the same vector also yields the retrieval embedding after $\ell_2$ normalization, any information required to reconstruct the counterpart text is forced to pass through the embedding itself, which is exactly the bottleneck we want to supervise.

\paragraph{Next Token Prediction (NTP).}
Consider the Query-to-Document (Q$\rightarrow$D) direction. Let the document-side text be $x_d=(x_d^1,\dots,x_d^T)$. We prepend the query prefix $\tilde{\mathbf{z}}_q$ and feed the concatenated sequence back into the same backbone for autoregressive decoding:
\begin{equation}
\mathbf{h}_t = F_\theta\!\left([\,\tilde{\mathbf{z}}_q\,;\,
x_d^{1},\dots,x_d^{t-1}\,]\right),
\qquad t=1,\dots,T,
\label{eq:s2b-decode-hidden}
\end{equation}
where $[\cdot\,;\cdot]$ denotes sequence concatenation and $\mathbf{h}_t$ is the decoder hidden state at step $t$. The next-token distribution is obtained by a softmax over the vocabulary $\mathcal{V}$:
\begin{equation}
P_\theta\!\left(x_d^{t}\mid \tilde{\mathbf{z}}_q, x_d^{<t}\right)
= \frac{\exp\!\left(\mathbf{e}_{x_d^{t}}^{\top}\mathbf{h}_t\right)}
{\sum_{v\in\mathcal{V}}\exp\!\left(\mathbf{e}_{v}^{\top}\mathbf{h}_t\right)},
\label{eq:s2b-softmax}
\end{equation}
where $\mathbf{e}_v$ is the embedding of token $v$. The NTP loss is the cross-entropy computed \emph{only} over document-side text tokens:
\begin{equation}
\mathcal{L}_{\text{NTP}}^{q\rightarrow d}
= -\frac{1}{T}\sum_{t=1}^{T}
\log P_\theta\!\left(x_d^{t}\mid \tilde{\mathbf{z}}_q, x_d^{<t}\right).
\label{eq:s2b-ntp-q2d}
\end{equation}
Symmetrically, the Document-to-Query (D$\rightarrow$Q) direction conditions on the document embedding $\tilde{\mathbf{z}}_d$ to reconstruct the query-side text $x_q=(x_q^1,\dots,x_q^{T'})$:
\begin{equation}
\mathcal{L}_{\text{NTP}}^{d\rightarrow q}
= -\frac{1}{T'}\sum_{t=1}^{T'}
\log P_\theta\!\left(x_q^{t}\mid \tilde{\mathbf{z}}_d, x_q^{<t}\right).
\label{eq:s2b-ntp-d2q}
\end{equation}

\paragraph{Multi-Token Prediction (MTP).}
NTP only supervises the immediate next token, so the embedding may encode short-range surface cues rather than the longer-range semantics needed for fine-grained matching. Following the multi-token prediction objective~\cite{MultiTokenPrediction,DeepSeekV3}, we extend the decoding supervision to $D$ additional future tokens at every position. Different from parallel-head designs, we adopt the \emph{sequential} formulation that keeps a complete causal chain at each prediction depth. We describe the Q$\rightarrow$D direction; the D$\rightarrow$Q direction is symmetric.

We attach $D$ lightweight MTP modules on top of the conditional decoder. The $k$-th module ($k=1,\dots,D$) consists of a Transformer block $\mathrm{TRM}_k$ and a projection $M_k\in\mathbb{R}^{h\times 2h}$, and it \emph{shares} the input embedding layer $\mathrm{Emb}(\cdot)$ and the output head $\mathrm{OutHead}(\cdot)$ with the main decoder. Let $\mathbf{h}_t^{0}\!=\!\mathbf{h}_t$ denote the main-decoder hidden state from Eq.~\eqref{eq:s2b-decode-hidden}. At depth $k$, the module combines the depth-$(k\!-\!1)$ representation with the embedding of the $(t\!+\!k)$-th document token:
\begin{equation}
\mathbf{h}_t^{\prime k}
= M_k\!\left[\,\mathrm{RMSNorm}\!\left(\mathbf{h}_t^{k-1}\right);\,
\mathrm{RMSNorm}\!\left(\mathrm{Emb}(x_d^{\,t+k})\right)\right],
\label{eq:s2b-mtp-combine}
\end{equation}
\begin{equation}
\mathbf{h}_{1:T-k}^{k} = \mathrm{TRM}_k\!\left(\mathbf{h}_{1:T-k}^{\prime k}\right),
\qquad
P_{t}^{k}
= \mathrm{OutHead}\!\left(\mathbf{h}_t^{k}\right),
\label{eq:s2b-mtp-head}
\end{equation}
where $[\cdot\,;\cdot]$ is concatenation and $P_{t}^{k}\in\mathbb{R}^{|\mathcal{V}|}$ is the predicted distribution for the $(t\!+\!k)$-th token. Each depth contributes a cross-entropy term, and the Q$\rightarrow$D MTP loss averages over the $D$ depths:
\begin{equation}
\mathcal{L}_{\text{MTP}}^{q\rightarrow d}
= \frac{1}{D}\sum_{k=1}^{D}
\left(-\frac{1}{T-k}\sum_{t=1}^{T-k}
\log P_{t}^{k}\!\left[x_d^{\,t+k}\right]\right),
\label{eq:s2b-mtp-q2d}
\end{equation}
where $P_{t}^{k}[\cdot]$ selects the probability of the ground-truth token. The D$\rightarrow$Q loss $\mathcal{L}_{\text{MTP}}^{d\rightarrow q}$ is defined analogously by conditioning on $\tilde{\mathbf{z}}_d$ and decoding the query-side text. Predicting several tokens ahead densifies the token-level signal and forces the embedding to pre-plan counterpart semantics beyond the immediate next token.

\paragraph{Training-only supervision, inference-time efficiency.}
Both NTP and MTP are pure training-time objectives. The decoding pass and the $D$ MTP modules are used only to back-propagate token-level gradients into the shared backbone and the readout; they are discarded at inference time. DME therefore remains a standard bi-encoder retriever: each query or document is encoded once into a dense vector, and retrieval is performed by vector similarity without any decoding, autoregressive generation, or cross-encoder scoring. As a result, cross-conditional reconstruction strengthens what the embedding must preserve while adding zero overhead to the online retrieval pipeline.

\paragraph{Stage 2-B objective.}
We combine the bidirectional NTP and MTP terms into the Stage-2B objective:
\begin{equation}
    \mathcal{L}_{\mathrm{S2B}}
    =
    \lambda_{\mathrm{NTP}}
    \left(
        \mathcal{L}_{\mathrm{NTP}}^{q\rightarrow d}
        +
        \mathcal{L}_{\mathrm{NTP}}^{d\rightarrow q}
    \right)
    +
    \lambda_{\mathrm{MTP}}
    \left(
        \mathcal{L}_{\mathrm{MTP}}^{q\rightarrow d}
        +
        \mathcal{L}_{\mathrm{MTP}}^{d\rightarrow q}
    \right).
\end{equation}
Jointly optimizing the two directions forces each embedding to encode enough counterpart-side semantics to reconstruct its match, which is exactly the semantic sufficiency that Stage 2-B is designed to supplement.

\subsection{Joint Training Objective}
\label{sec:joint-training-objective}

DME is optimized sequentially. Stage 1 uses the large-scale contrastive objective $\mathcal{L}_{\mathrm{S1}}^{\mathrm{CL}}$ defined in Eq.~\eqref{eq:stage1-cl}. During Stage 2, we continue contrastive retrieval training on the Stage-2 data mixture and denote this term as $\mathcal{L}_{\mathrm{S2}}^{\mathrm{CL}}$. It has the same form as the Stage-1 contrastive objective but is computed using the Stage-2 data and readout representations.

The Stage-2 objective is
\begin{equation}
    \mathcal{L}_{\mathrm{S2}}
    =
    \mathcal{L}_{\mathrm{S2}}^{\mathrm{CL}}
    +
    \mathcal{L}_{\mathrm{S2A}}
    +
    \mathcal{L}_{\mathrm{S2B}}.
\end{equation}
The contrastive term maintains the global retrieval geometry, Stage 2-A provides evidence and latent-state supervision, and Stage 2-B encourages the readout representation to preserve counterpart-side textual semantics.

\section{Experiments and Analysis}
\label{sec:experiments}

\subsection{Experimental Setup}
\label{sec:experimental-setup}

\paragraph{Evaluated models.}
We evaluate DME at two model scales, denoted as DME-2B and DME-9B. Both models follow the same two-stage training framework described in Section~\ref{sec:multi-stage-learning-framework}. Stage 1 performs large-scale contrastive pre-training, while Stage 2 further applies Evidence-Grounded Typed Latent Reasoning and Cross-Conditional Reconstruction. Unless otherwise specified, the final DME models use the complete Stage-2 training recipe.

\paragraph{Training configuration.}
We use LoRA-based parameter-efficient fine-tuning for both DME-2B and DME-9B, and tune key contrastive learning hyperparameters such as batch size, learning rate, and temperature for stable large-batch training. To control the cost of multimodal inputs, we limit the image token budget to 1280 tokens and uniformly sample 32 frames per video. Training is further stabilized with BF16 training, gradient checkpointing, and ZeRO optimization, which reduce memory consumption and enable larger batches under long multimodal inputs. Section~\ref{sec:training-recipe-diagnostics} further analyzes the effects of batch size, in-batch sampling, image token budget, and video frame sampling.

\paragraph{Benchmarks.}
For public evaluation, we report results on MMEB-v2, a comprehensive benchmark for universal multimodal embedding that covers diverse retrieval tasks across text, image, video, visual document, and mixed-modality inputs~\cite{MMEB,VLM2VecV2,Qwen3VLEmbedding}. We use MMEB-v2 as the main benchmark to compare DME against existing multimodal embedding models under comparable model-scale regimes. For internal evaluation, we additionally use Douyin's in-house offline evaluation set, which reflects real-world production retrieval traffic across cross-modal directions, to assess whether the DME techniques transfer to a large-scale industrial setting.

\paragraph{Baselines.}
We compare DME with representative MLLM-based multimodal embedding models and recent reasoning-enhanced embedding methods. The MLLM-based embedding baselines include GME~\cite{GME}, VLM2Vec/VLM2Vec-V2~\cite{MMEB,VLM2VecV2}, and Qwen3-VL-Embedding~\cite{Qwen3VLEmbedding}. We also compare with recent reasoning-oriented retrieval representation methods, including Think-Then-Embed (TTE)~\cite{ThinkThenEmbed}, TTE-v2~\cite{ThinkThenEmbedV2}, and Embed-RL~\cite{EmbedRL}. When reporting scale-wise comparisons, we group baselines by model size to ensure that DME-2B and DME-9B are compared with models of comparable capacity.

\paragraph{Metrics.}
Following the standard MMEB-v2 evaluation protocol, we report the official aggregate score together with task-level or modality-level breakdowns when available. For retrieval tasks, we use ranking-based metrics such as Recall@$K$, nDCG@$K$, or task-specific official metrics depending on the benchmark definition. For internal Douyin evaluation, we report the relative improvement of DME over the in-production baseline on Douyin's in-house offline evaluation set.

\begin{table*}[t]
\centering
\small
\setlength{\tabcolsep}{3.1pt}
\renewcommand{\arraystretch}{1.08}
\resizebox{\linewidth}{!}{
\begin{tabular}{l c ccccc ccccc ccccc c}
\toprule
\multirow{2}{*}{\textbf{Model}} &
\multirow{2}{*}{\textbf{Size}} &
\multicolumn{5}{c}{\textbf{Image}} &
\multicolumn{5}{c}{\textbf{Video}} &
\multicolumn{5}{c}{\textbf{VisDoc}} &
\multirow{2}{*}{\textbf{All}} \\
\cmidrule(lr){3-7}
\cmidrule(lr){8-12}
\cmidrule(lr){13-17}
& & CLS & QA & RET & GD & Avg.
& CLS & QA & RET & MRET & Avg.
& VDRv1 & VDRv2 & VR & OOD & Avg.
& \\
\midrule
\# datasets
& -- & 10 & 10 & 12 & 4 & 36
& 5 & 5 & 5 & 3 & 18
& 10 & 4 & 6 & 4 & 24
& 78 \\
\midrule
VLM2Vec~\cite{MMEB} & 8B
& 62.7 & 56.9 & 69.4 & 82.2 & 65.5
& 39.1 & 30.0 & 29.0 & 38.9 & 33.7
& 56.9 & 9.4 & 59.1 & 54.0 & 49.1
& 53.1 \\
VLM2Vec-V2~\cite{VLM2VecV2} & 2B
& 62.9 & 56.3 & 69.5 & 77.3 & 64.9
& 39.3 & 34.3 & 28.8 & 36.8 & 34.6
& 75.5 & 44.9 & 79.4 & 62.2 & 69.2
& 59.2 \\
GME~\cite{GME} & 8B
& 57.7 & 34.7 & 71.2 & 59.3 & 56.0
& 37.4 & 50.4 & 28.4 & 37.0 & 38.4
& 89.4 & 55.6 & 85.0 & 68.3 & 79.3
& 59.1 \\
Ops-MM-Embedding-v1\modelhome{https://huggingface.co/OpenSearch-AI/Ops-MM-embedding-v1-7B} & 8B
& 69.7 & 69.6 & 73.1 & 87.2 & 72.7
& 59.7 & 62.2 & 45.7 & 43.2 & 53.8
& 80.1 & 59.6 & 79.3 & 67.8 & 74.4
& 68.9 \\
RzenEmbed~\cite{RzenEmbed}\modelhome{https://github.com/360CVGroup/RzenEmbed} & 8B
& 70.6 & 71.7 & 78.5 & 92.1 & 75.9
& 58.8 & 63.5 & 51.0 & 45.5 & 55.7
& 89.7 & 60.7 & 88.7 & 69.9 & 81.3
& 72.9 \\

Embed-RL~\cite{EmbedRL} & 4B
& 63.7 & 70.5 & 71.3 & 91.4 & 70.1
& 57.6 & 58.4 & 45.1 & 49.5 & 53.0
& 80.2 & 53.4 & 84.9 & 67.1 & 74.7
& 68.1 \\

WeMM-Embedding & 2B
& 72.1 & 72.6 & 76.6 & 93.3 & 76.1
& 61.5 & 64.0 & 54.2 & 52.6 & 58.7
& 87.2 & 53.4 & 88.8 & 34.5 & 73.2
& 71.2 \\

Qwen3-VL-Embedding~\cite{Qwen3VLEmbedding}\modelhome{https://huggingface.co/Qwen/Qwen3-VL-Embedding-2B} & 2B
& 70.3 & 74.3 & 74.8 & 88.5 & 75.0
& 71.9 & 64.9 & 53.9 & 53.3 & 61.9
& 84.4 & 65.3 & 86.4 & 69.4 & 79.2
& 73.2 \\
IFM-TTE~\cite{ThinkThenEmbed}\modelhome{https://interestfm-tte.github.io/} & 8B
& 76.7 & 78.5 & 74.6 & 89.3 & 77.9
& 60.5 & 67.9 & 51.7 & 54.9 & 59.2
& 85.2 & 71.5 & 92.7 & 53.3 & 79.5
& 74.1 \\

TTE-v2$^{*}$~\cite{ThinkThenEmbedV2} & 7B
& 78.1 & 79.0 & 76.3 & 91.6 & 79.2
& 58.3 & 66.9 & 52.3 & 68.0 & 60.7
& 85.4 & 63.3 & 94.3 & 69.0 & 82.3
& 75.7 \\

WeMM-Embedding & 8B
& 73.6 & 76.1 & 78.6 & 92.9 & 78.1
& 66.5 & 71.7 & 56.4 & 55.2 & 63.2
& 89.5 & 59.3 & 90.4 & 35.1 & 75.6
& 73.9 \\

Qwen3-VL-Embedding~\cite{Qwen3VLEmbedding}\modelhome{https://huggingface.co/Qwen/Qwen3-VL-Embedding-8B} & 8B
& 74.2 & 81.1 & 80.2 & 92.3 & 80.1
& 78.4 & 71.0 & 58.7 & 56.1 & 67.1
& 87.2 & 69.9 & 88.7 & 73.3 & 82.4
& 77.8 \\
\midrule
\textbf{DME} & \textbf{2B}
& \textbf{68.6} & \textbf{76.2} & \textbf{75.6} & \textbf{94.3} & \textbf{75.9}
& \textbf{84.5} & \textbf{61.9} & \textbf{55.5} & \textbf{57.4} & \textbf{65.6}
& \textbf{87.0} & \textbf{56.2} & \textbf{90.3} & \textbf{70.0} & \textbf{79.9}
& \textbf{74.8} \\
\textbf{DME} & \textbf{9B}
& \textbf{74.5} & \textbf{80.9} & \textbf{78.2} & \textbf{94.6} & \textbf{79.8}
& \textbf{87.7} & \textbf{71.0} & \textbf{61.0} & \textbf{58.5} & \textbf{70.8}
& \textbf{87.6} & \textbf{57.8} & \textbf{94.5} & \textbf{73.5} & \textbf{82.0}
& \textbf{78.4} \\
\bottomrule
\end{tabular}
}
\caption{
\textbf{MMEB-v2 evaluation with task-family breakdown.} CLS, QA, RET, GD, MRET, VDR, VR, and OOD denote classification, question answering, retrieval, grounding, moment retrieval, ViDoRe, VisRAG, and out-of-distribution evaluation, respectively. Unless otherwise noted, all reported scores are taken from the official MMEB Leaderboard. The $\dagger$ marker indicates an available checkpoint, model card, or project homepage. $^{*}$TTE-v2 is reported from its paper under a 76-task MMEB-v2 setting that excludes two VisDoc OOD datasets; other rows follow the 78-task setting when available.
}
\label{tab:mmeb-v2-results}
\end{table*}

\subsection{Results on MMEB-v2}
\label{sec:mmeb-v2-results}

Table~\ref{tab:mmeb-v2-results} reports the main results on MMEB-v2. We compare DME with recent MLLM-based multimodal embedding models and reasoning-enhanced retrieval models under comparable scale regimes. DME-2B obtains an overall score of \textbf{74.8}, and DME-9B obtains an overall score of \textbf{78.4}. From the results, we make three observations.

\paragraph{(1) DME achieves strong scale-wise performance.}
DME-2B should be primarily compared with other 2B-level models such as VLM2Vec-V2 and Qwen3-VL-Embedding-2B, while DME-9B should be compared with larger models such as RzenEmbed-8B, IFM-TTE-8B, TTE-v2-7B, and Qwen3-VL-Embedding-8B. Under this comparison, DME shows strong performance at both model scales, indicating that the proposed two-stage training framework is effective for both compact and larger MLLM-based embedders.

\paragraph{(2) The gains are not concentrated in a single modality group.}
DME performs consistently across Image, Video, and VisDoc groups. For the 2B and 9B variants, DME obtains \textbf{75.9}/\textbf{79.8} on Image, \textbf{65.6}/\textbf{70.8} on Video, and \textbf{79.9}/\textbf{82.0} on VisDoc, respectively. This broad improvement is important because MMEB-v2 contains heterogeneous task families, including classification, question answering, retrieval, grounding, moment retrieval, visual-document retrieval, VisRAG, and out-of-distribution evaluation.

\paragraph{(3) DME outperforms reasoning-enhanced baselines while using lightweight latent-token inference.}
Recent models such as IFM-TTE, TTE-v2, and Embed-RL introduce reasoning signals to improve multimodal retrieval representations. DME follows a different design: evidence grounding and typed latent reasoning are integrated into the encoder through a small number of retrieval-specific latent tokens, instead of generating explicit chain-of-thought text or applying cross-encoder reranking at evaluation time. The results in Table~\ref{tab:mmeb-v2-results} suggest that this latent-token design provides strong retrieval quality while keeping inference close to the standard dense retrieval pipeline.

\label{sec:douyin-industrial-benchmark}

\begin{table}[t]
\centering
\begin{tabular}{lccccc}
\toprule
 & Text2Video & Text2Image & Image2Image & Image2Video & All \\
\midrule
$\Delta$ & +3.10\% & +3.03\% & +2.70\% & +2.83\% & +2.92\% \\
\bottomrule
\end{tabular}
\caption{\textbf{Relative performance gains on Douyin industrial benchmark.} We initialize the in-house model from DME and continue
training with several DME techniques transferred to the in-house setting.
Improvements ($\Delta$) are relative gains over the previous production model.}
\label{tab:inhouse-transfer}
\end{table}

\subsection{Industrial Deployment}
\label{sec:industrial-deployment}

Beyond public benchmarks, we further validate DME in Douyin's production retrieval system through both offline and online evaluations.

\textbf{Offline results.}
We initialize the in-house model from DME and continue training with several DME techniques transferred to this setting. As shown in Table~\ref{tab:inhouse-transfer}, this yields a 2.92\% relative improvement in overall retrieval quality over the previous production model. The gains are consistent across all four directions, ranging from +3.10\% on Text2Video to +2.70\% on Image2Image, which indicates that the DME techniques generalize to a large-scale industrial retrieval setting rather than being effective only under public academic benchmarks.

\textbf{Online results.}
We further deploy the DME-based model in Douyin's online search system, powering scenarios such as generative search and serving as a retrieval feature for ranking. Online A/B testing on Douyin search verifies a 0.1\% Lifetime (LT) gain in core online business metrics.

\subsection{Ablation Study}
\label{sec:training-recipe-diagnostics}

\textbf{Training Recipe Ablation Study.} To inspect how each part of the DME learning recipe affects retrieval quality, we analyze the cumulative DME training recipe on MMEB-v2. All variants are trained on the same Stage-2 data and differ only in which components of the recipe are enabled, so the comparison isolates the marginal effect of each component rather than removing modules from the final model. Starting from a model that is trained directly on this data, we progressively enable large-scale Stage-1 pre-training, Stage 2-A, and Stage 2-B, and report how the model changes across all task groups.

We report four configurations. Baseline is initialized from Qwen3.5 and trained directly on the Stage-2 data with the contrastive objective only, without large-scale Stage-1 pre-training and without either Stage-2 mechanism. +Stage 1 adds large-scale contrastive pre-training prior to this training. +Stage 2-A further adds Evidence-Grounded Typed Latent Reasoning, including anchor-based evidence grounding, typed query-side latent supervision,
and evidence-enhanced readout. +Stage 2-B adds Cross-Conditional Reconstruction with NTP/MTP, giving the final DME model.

\begin{table*}[t]
\centering
\setlength{\tabcolsep}{12pt}
\label{tab:ablation}
\begin{tabular}{ccccccc}
\toprule
\multicolumn{3}{c}{Configuration} & \multicolumn{4}{c}{Score} \\
\cmidrule(lr){1-3} \cmidrule(lr){4-7}
Stage 1 & Stage 2-A & Stage 2-B & Image & Video & VisDoc & All \\
\midrule
           &            &            & 74.6 & 55.3 & 77.1 & 70.9 \\
\checkmark &            &            & 74.8 & 59.3 & 79.0 & 72.5 \\
\checkmark & \checkmark &            & 75.2 & 63.7 & 79.2 & 73.8 \\
\checkmark & \checkmark & \checkmark & \textbf{75.9} & \textbf{65.6} & \textbf{79.9} & \textbf{74.8} \\
\bottomrule
\end{tabular}
\caption{
\textbf{Cumulative analysis of the DME training recipe on MMEB-v2.}
All configurations use the Stage-2 contrastive objective; checkmarks indicate
additional training components enabled on top of this common baseline.
Stage 2-A denotes Evidence-Grounded Typed Latent Reasoning, and Stage 2-B denotes
Cross-Conditional Reconstruction. Scores are reported on the Image, Video, and
VisDoc groups together with the overall average.
}
\label{tab:ablation}
\end{table*}

Table~\ref{tab:ablation} summarizes the cumulative effect of the DME recipe, which raises the overall score from 70.9 to 74.8, a gain of 3.9 points. Adding Stage 1 improves the overall score from 70.9 to 72.5, confirming that large-scale heterogeneous contrastive pre-training establishes a stronger unified embedding space even when the same Stage-2 data is used; the gain is concentrated on video (55.3 to 59.3) and visual-document retrieval (77.1 to 79.0), while image retrieval is largely unchanged (74.6 to 74.8). Adding Stage 2-A contributes 1.3 points overall (72.5 to 73.8) and yields the largest single improvement on video (59.3 to 63.7), consistent with its design of grounding retrieval in localized textual, visual, and temporal evidence. Adding Stage 2-B brings the overall score to 74.8, with balanced gains across all three groups (image 75.2 to 75.9, video 63.7 to 65.6, visual-document 79.2 to 79.9), indicating that preserving fine-grained counterpart-side semantics through cross-conditional reconstruction benefits retrieval broadly rather than any single modality.

\textbf{Training Parameter Ablation Study.} Beyond the main training recipe, we also identify several practical factors that have a substantial impact on multimodal contrastive training. These factors are not specific to a single DME module, but provide reusable guidance for training large-scale multimodal embedding models. We summarize three findings: batch negative space scaling, batch mixed sampling, and visual budget.

\paragraph{Scaling the in-batch negative space.}

Increasing the batch size expands the number of in-batch negatives, thereby making contrastive learning more challenging and improving the model's discriminative capability. We observe that increasing the batch size from 128 to 8192 brings consistent improvements across evaluation metrics. However, further increasing the batch size leads to diminishing returns and may even slightly degrade performance. This is likely because larger batches also increase the probability of false negatives within the batch, while the gradients from truly hard negatives can be diluted by a large number of easy negatives. These results suggest that simply scaling batch size is beneficial only up to a certain point, and should be combined with false-negative filtering and hard-negative construction.

\begin{wrapfigure}{r}{0.5\linewidth} 
    \vspace{-15pt} 
    \centering
    \includegraphics[width=\linewidth]{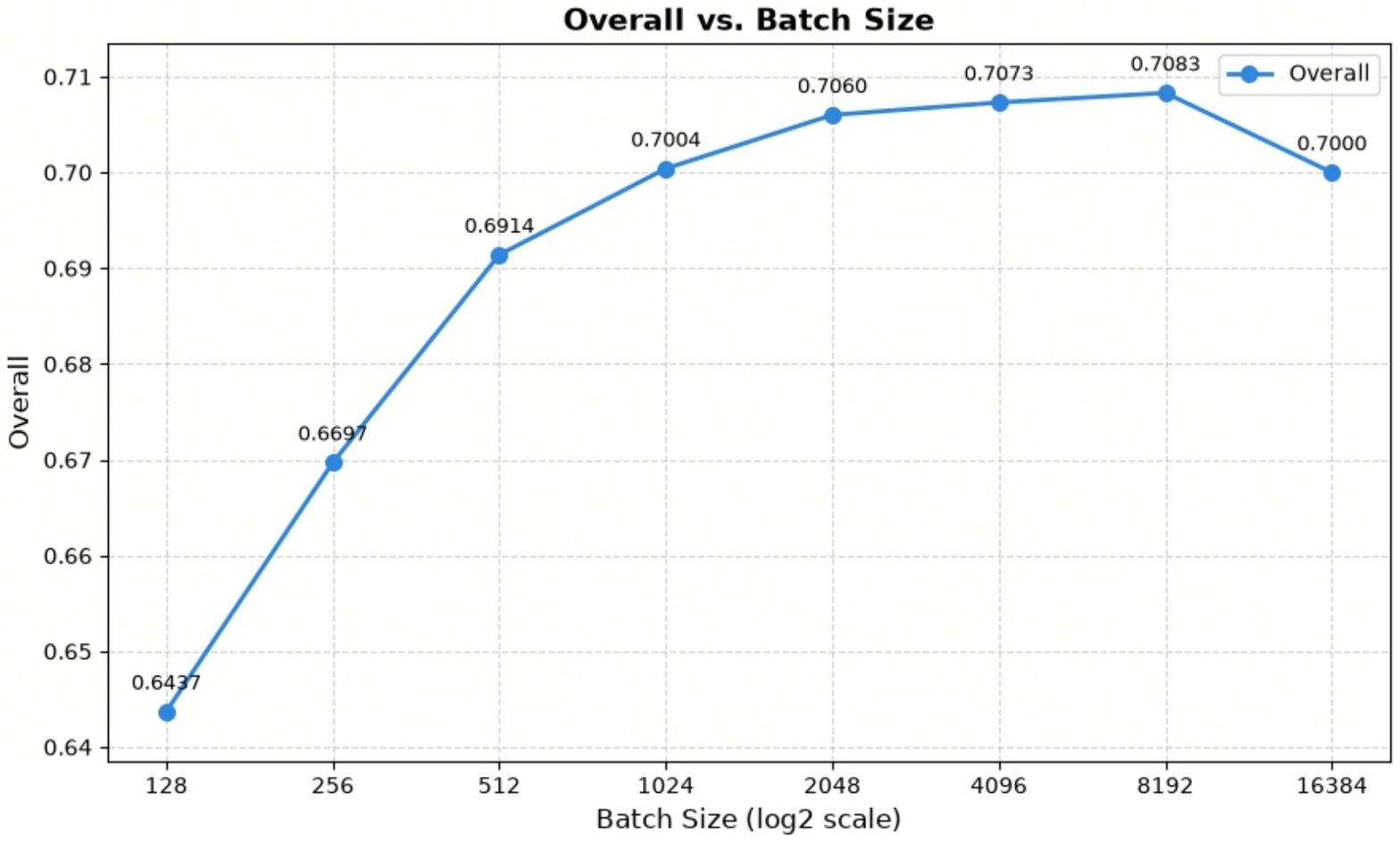}
\caption{\textbf{Effect of batch size on contrastive training.} Increasing batch size expands the in-batch negative space and improves retrieval performance up to a saturation point.}
    \label{fig:loss_similarity}
    \vspace{-15pt} 
\end{wrapfigure}

\paragraph{Task-balanced in-batch mixed sampling.}

We further study task-balanced in-batch mixed sampling, where samples from different data sources are mixed within the same batch according to a controlled ratio. As shown in Table~\ref{tab:mixed-sampling}, a mixing ratio of 0.25 achieves the best overall score. When the ratio is too small, the model benefits less from cross-task and cross-modal contrastive signals. When the ratio becomes too large, the batch may contain overly heterogeneous samples, weakening task-level consistency. Therefore, moderate in-batch mixing provides a better balance between stable same-source training and diverse cross-task supervision.

\begin{table}[t]
\centering
\small
\begin{tabular}{ccccc}
\toprule
Mixing Ratio & Overall & Image-Hit@1 & Video-Hit@1 & Doc-NDCG@5 \\
\midrule
0    & 0.6727 & 0.6914 & 0.5015 & 0.7734 \\
0.25 & \textbf{0.6919} & 0.7209 & \textbf{0.5202} & \textbf{0.7772} \\
0.5  & 0.6892 & \textbf{0.7266} & 0.5121 & 0.7659 \\
1    & 0.6861 & 0.7260 & 0.5106 & 0.7579 \\
\bottomrule
\end{tabular}
\caption{\textbf{Effect of task-balanced in-batch mixed sampling.} A mixing ratio of 0.25 achieves the best overall performance, balancing same-source consistency and cross-task diversity.}
\label{tab:mixed-sampling}
\end{table}

\paragraph{Optimizing visual resolution and video frame sampling.}

The visual input budget also has a clear impact on retrieval quality. For videos, increasing the number of sampled frames improves the model's ability to capture long-range actions and temporal semantics. As shown in Table~\ref{tab:video-frame-sampling}, increasing inference frames from 8 to 32 improves Video-Hit@1 under the 8-frame training setting, while training with 32 frames further improves the best performance. The gain from increasing inference frames beyond 32 is relatively small, suggesting that 32 frames provide a good trade-off between temporal coverage and efficiency.

For images, increasing the token budget improves the model's ability to preserve fine-grained visual details. As shown in Table~\ref{tab:image-token-budget}, increasing the image token budget from 256 to 1280 brings consistent gains on Overall, Image-Hit@1, Video-Hit@1, and Doc-NDCG@5. Based on these results, we use 1280 image tokens and 32 video frames in the final training configuration, which balances retrieval quality and computational efficiency.

\begin{table}[t]
\centering
\small
\begin{tabular}{ccc}
\toprule
Training Frames & Inference Frames & Video-Hit@1 \\
\midrule
8  & 8  & 0.5429 \\
8  & 32 & 0.5705 \\
8  & 64 & 0.5660 \\
\midrule
32 & 8  & 0.5346 \\
32 & 32 & 0.5910 \\
32 & 64 & \textbf{0.5920} \\
\bottomrule
\end{tabular}
\caption{\textbf{Effect of video frame sampling.} Increasing the number of sampled frames improves video retrieval, with 32 training frames providing a strong trade-off between temporal coverage and computational cost.}
\label{tab:video-frame-sampling}
\end{table}

\begin{table}[t]
\centering
\small
\begin{tabular}{ccccc}
\toprule
Image Tokens & Overall & Image-Hit@1 & Video-Hit@1 & Doc-NDCG@5 \\
\midrule
256  & 0.7049 & 0.7441 & 0.5429 & 0.7675 \\
1280 & \textbf{0.7090} & \textbf{0.7457} & \textbf{0.5527} & \textbf{0.7713} \\
\bottomrule
\end{tabular}
\caption{\textbf{Effect of image token budget.} Increasing the image token budget improves visual detail preservation and leads to consistent gains across evaluation metrics.}
\label{tab:image-token-budget}
\end{table}

\subsection{Measuring Semantic Sufficiency via Representation Completeness}
\label{sec:ntp-topk}
To quantify how much content a single embedding actually preserves, we recast
the NTP objective of Eq.~\eqref{eq:s2b-ntp-q2d} as a bounded, interpretable
metric. Instead of the unbounded cross-entropy, we ask whether the trained DME
model can recover each target token from the pooled embedding alone. Consider
the Q$\rightarrow$D direction. We prepend the query prefix
$\tilde{\mathbf{z}}_q$ as the \emph{only} conditioning input, and run a single
teacher-forcing pass through the same backbone over the document-side 
text $x_d=(x_d^1,\dots,x_d^{t})$. At each position, this yields the         
next-token distribution
$P_\theta\!\left(x_d^{t}\mid \tilde{\mathbf{z}}_q, x_d^{<t}\right)$
of Eq.~\eqref{eq:s2b-softmax}. A position counts as a hit if the ground-truth
token lies within the model's Top-$K$ predictions, and we report the
micro-averaged accuracy over the first $L_i$ evaluated positions of each sample:
\begin{equation}
\mathrm{acc}@K \;=\;
\frac{\displaystyle\sum_{i}\sum_{t=1}^{L_i}
      \mathbf{1}\!\left[\,x^{t,(i)}\in
      \mathrm{Top}_K\!\big(P_\theta(\cdot\mid \tilde{\mathbf{z}}^{(i)}, x^{<t,(i)})\big)\right]}
     {\displaystyle\sum_{i} L_i}\;\in\;[0,1], 
\label{eq:acck}
\end{equation}
where $T_i$ is the target length of sample $i$, $L_{\max}$ is a fixed truncation length so that only the first $L_i=\min(L_{\max}, T_i)$ positions of each sample are evaluated (we set $L_{\max}=10$), and $\mathrm{Top}_K(\cdot)$ is the set of $K$ tokens with the largest probability. Unlike the NTP cross-entropy, $\mathrm{acc}@K$ lies in $[0,1]$ and reads directly as the fraction of the first $L_i$ target tokens the embedding can recover within its Top-$K$ guesses, making it comparable across datasets, directions, and
modalities.

We probe four prefix$\rightarrow$target directions per retrieval pair, reusing
the training-time branch construction. The self directions, q2q and d2d,
condition on $\tilde{\mathbf{z}}_q$ (resp.\ $\tilde{\mathbf{z}}_d$) to recover
its own text $x_q$ (resp.\ $x_d$), and thus measure \emph{reconstruction
completeness}: how faithfully an embedding re-encodes its own content. The
cross directions, q2d and d2q, condition on one side to recover the other
(mirroring $\mathcal{L}_{\text{NTP}}^{q\rightarrow d}$ and
$\mathcal{L}_{\text{NTP}}^{d\rightarrow q}$), and thus measure
\emph{counterpart completeness}: how much of the paired document (or query) a
query (or document) embedding retains. A high $\mathrm{acc}@K$ indicates that
the pooled vector exposes, through the DME model, the token-level content
needed to regenerate the target, evidence that the representation is
information-complete rather than merely discriminative for retrieval. This
directly operationalizes the notion of semantic sufficiency that motivates DME.

\begin{table}[t]
\centering
\resizebox{\linewidth}{!}{%
\begin{tabular}{l ccc ccc ccc ccc}
\toprule
\multirow{2}{*}{Direction} & \multicolumn{3}{c}{Image} & \multicolumn{3}{c}{Video} & \multicolumn{3}{c}{VisDoc} & \multicolumn{3}{c}{All} \\
\cmidrule(lr){2-4}\cmidrule(lr){5-7}\cmidrule(lr){8-10}\cmidrule(lr){11-13}
 & @1 & @5 & @10 & @1 & @5 & @10 & @1 & @5 & @10 & @1 & @5 & @10 \\
\midrule
\multicolumn{13}{l}{\emph{Self (reconstruction completeness)}} \\
q2q & 0.8758 & 0.9469 & 0.9542 & 0.8781 & 0.8984 & 0.9072 & 0.8909 & 0.9041 & 0.9102 & 0.8792 & 0.9184 & 0.9262 \\
d2d & 0.7737 & 0.9117 & 0.9387 & 0.6431 & 0.8640 & 0.9136 & 0.9455 & 0.9805 & 0.9903 & 0.7432 & 0.9014 & 0.9355 \\
\midrule
\multicolumn{13}{l}{\emph{Cross (counterpart completeness)}} \\
q2d & 0.6279 & 0.8102 & 0.8568 & 0.5920 & 0.8328 & 0.8787 & 0.9513 & 0.9925 & 0.9987 & 0.6585 & 0.8453 & 0.8859 \\
d2q & 0.8736 & 0.9478 & 0.9551 & 0.8728 & 0.8989 & 0.9104 & 0.8978 & 0.9200 & 0.9337 & 0.8771 & 0.9216 & 0.9318 \\
\bottomrule
\end{tabular}}
\caption{\textbf{Representation completeness measured by teacher-forced Top-$K$
accuracy (acc@$K$).} Given a pooled, un-normalized embedding as the only
conditioning prefix, the trained DME model is asked to recover the target
tokens under teacher forcing; acc@$K$ is the fraction of evaluated positions
whose ground-truth token falls within the model's Top-$K$ predictions. Scores
are micro-averaged over the first $L_i$ evaluated positions of each sample ($\sum\text{hits}/\sum\text{positions}$),
and ``All'' merges the three domains with token-level weighting. Self directions
(q2q, d2d) probe how faithfully an embedding re-encodes its own text; cross
directions (q2d, d2q) probe how much of the paired counterpart it retains.}
\label{tab:ntp-topk-micro}
\end{table}

Table~\ref{tab:ntp-topk-micro} reports representation completeness across the
four directions and three domains. Overall, DME embeddings are highly
information-complete: on the merged set, the self directions recover the
ground-truth token at Top-1 in $87.9\%$ (q2q) and $74.3\%$ (d2d) of positions,
rising above $92\%$ at Top-10. This confirms that a single pooled vector
retains most of the token-level content of its own input rather than collapsing
to a coarse, retrieval-only summary. The cross directions are naturally harder,
since reproducing the counterpart requires information that is only shared
through the matched pair; nonetheless, d2q reaches $87.7\%$ at Top-1 and q2d
reaches $65.9\%$, indicating that a substantial fraction of the paired content
is genuinely encoded, consistent with the counterpart-side supervision used
during Stage-2 training. Across domains, VisDoc is the easiest to reconstruct
(q2d reaches $95.1\%$ at Top-1), matching its text-heavy, highly regular
content, whereas video is the hardest in the cross setting (q2d $59.2\%$ at
Top-1), reflecting the larger information gap between a text query and a
temporally extended visual target. Taken together, these results provide a
direct, quantitative confirmation that DME representations are semantically
sufficient, and the metric offers an interpretable signal that we further use
to guide representation optimization in industrial settings.

\subsection{Efficiency and Query Latency}
\label{sec:efficiency-query-latency}

A key design goal of DME is to introduce evidence-grounded latent reasoning without sacrificing the serving efficiency required by large-scale retrieval systems such as Douyin. Unlike explicit CoT generation or reranking-based reasoning, Stage 2-A only inserts a small number of anchor and typed latent tokens into the query encoder forward pass. We therefore measure the additional query-side latency introduced by these latent tokens.

The measurement only covers the forward pass of the query encoder. It excludes candidate encoding, query--candidate scoring, ANN retrieval, data loading, collation, and host-to-device transfer. 
We compare the same batches under two settings: \textbf{w/o Latent Tokens}, where the query uses the standard readout, and \textbf{w/ Latent Tokens}, where anchor and typed latent reasoning tokens are enabled. We use 8 high-performance GPUs with batch size 4. We first run 20 warmup iterations, which are not included in the timing statistics, and then record up to 200 measured iterations. Each iteration corresponds to one query-batch forward pass. We report p50 latency, \ie, the median latency over measured batches, which is a standard percentile metric for serving-time latency analysis.

\begin{table}[t]
\centering
\small
\setlength{\tabcolsep}{5pt}
\renewcommand{\arraystretch}{1.08}
\begin{tabular}{llcccc}
\toprule
\textbf{Query Type} & \textbf{Dataset}
& \shortstack{\textbf{w/o Latent}\\\textbf{p50 (ms/batch)}}
& \shortstack{\textbf{w/ Latent}\\\textbf{p50 (ms/batch)}}
& \shortstack{\textbf{$\Delta$}\\\textbf{(ms/batch)}}
& \shortstack{\textbf{$\Delta$}\\\textbf{(ms/query)}} \\
\midrule
Text & MSCOCO-T2I
& 41.962 & 45.450 & +3.487 & +0.87 \\
Image+Text & OK-VQA
& 111.084 & 111.399 & +0.315 & +0.08 \\
Video+Text & Video-MME
& 1049.479 & 1052.694 & +3.215 & +0.80 \\
\bottomrule
\end{tabular}
\caption{
\textbf{Query encoder p50 latency with and without latent reasoning tokens.} P50 denotes the median latency over measured batches. Per-query overhead is computed using batch size 4.
}
\label{tab:query-latency}
\end{table}

As shown in Table~\ref{tab:query-latency}, enabling latent reasoning introduces only minor overhead. The additional cost is below 1 ms per query for text and video queries, and nearly negligible for image--text queries. This is consistent with the design of DME: latent reasoning does not perform autoregressive generation or multi-round inference, but only adds a few soft tokens to the same encoder forward pass. For multimodal inputs, especially video, the computation is dominated by the original visual tokens, making the relative cost of the additional latent tokens small.

These results show that DME can keep latent reasoning active during query encoding while remaining close to the standard dense retrieval pipeline, which is important for practical deployment in large-scale multimodal retrieval.

\section{Conclusion}
\label{sec:conclusion}

In this report, we presented Douyin Multimodal Embedding (DME), a representation model for large-scale industrial multimodal retrieval. Through a two-stage training framework, DME first performs large-scale contrastive pre-training to establish a unified multimodal embedding space, and then supplements semantic sufficiency in Stage 2 via Evidence-Grounded Typed Latent Reasoning and Cross-Conditional Reconstruction, which respectively ground the embedding in retrieval-relevant evidence and enforce the preservation of fine-grained counterpart-side semantics. Cross-Conditional Reconstruction is used only during training, whereas Stage 2-A
retains a small number of latent tokens within the same encoder forward pass.
DME therefore remains a dense bi-encoder retriever, with only marginal
query-side latency overhead. In our experiments, DME achieves state-of-the-art results on MMEB-v2 among models of comparable scale, and our latency analysis confirms that the latent reasoning tokens introduce only marginal query-encoding overhead. We further quantify semantic sufficiency through a representation completeness measure, showing that counterpart-side content can be recovered from the learned embeddings at high token-level accuracy, which together with the latency analysis demonstrates that DME attains strong advantages in both semantic sufficiency and efficiency. Moreover, DME has been deployed in Douyin multimodal search, where it delivers significant gains across downstream businesses.

Looking forward, we are extending DME along two scaling directions to further strengthen the generality of the learned representations. The first is data scaling, where we enlarge and diversify the multimodal training corpus to broaden modality and task coverage. The second is model-size scaling, where we train DME with larger backbones to increase representational capacity. Our preliminary attempts along both directions already yield consistent gains, and we regard scaling data and model size as a promising path toward more general-purpose multimodal embeddings.

\newpage

\section{Contributor List}

\contributionlist

\textbf{Contributors:} \\
Haonan Chen$^{2,*,\dagger}$ \quad Chu Li$^{1,*}$ \quad Zhicheng Wang$^{1,*}$ \quad Yuanwei Liu$^{1}$ \quad Yuanjiang Wang$^{1}$

\textbf{Project Leader:} \\
Shaohua Jiang$^{1}$

\textbf{Supervisor:} \\
Zhicheng Dou$^{2}$

\textbf{Affiliations:} \\
$^{1}$ ByteDance Douyin Search Multimodal Team \\
$^{2}$ Gaoling School of Artificial Intelligence, Renmin University of China

\vspace{2pt}

\noindent
$*$ Equal contribution. Contributors are listed in alphabetical order by last name initial.

\noindent
$^\dagger$ Work was done during Haonan's internship at ByteDance Douyin Search Multimodal Team.

\section{Acknowledgments}
We sincerely thank Leyang Wang, Lanqing Hu, Tianlong Ma, Jianfeng Li, Xiangyuan Ren and Shikang Wu for their valuable support and contributions.

\clearpage

\bibliographystyle{plainnat}
\bibliography{main}

\appendix

\clearpage

\section*{Appendix}

\section{Detailed Formulation of Stage 2-A}
\label{app:stage2a-details}

This appendix provides the implementation-level formulation of the evidence assignment, assignment balancing, summary-conditioned supervision, and typed latent objectives summarized in Section~\ref{sec:stage2a-typed-latent-reasoning}.

\subsection{Anchor Evidence Assignment}
\label{app:anchor-evidence-assignment}

For training sample $i$, let
\begin{equation}
    \mathcal{S}_{\mathrm{anc}}
    =
    \{q,+\}
\end{equation}
denote the sides used for anchor-grounding supervision. Negative-side structured evidence is used to construct rejection-oriented trajectory states, but is not used as an anchor hit target.

For side $s \in \mathcal{S}_{\mathrm{anc}}$ and modality $m$, let $\mathcal{I}_m(i,s)$ denote the corresponding raw content-token indices, and let $R_m$ denote the number of anchor tokens assigned to modality $m$. Given anchor hidden state $\mathbf{a}_{i,s,r}^{(m)}$ and content-token hidden state $\mathbf{x}_{i,s,j}$, DME computes
\begin{equation}
    \ell_{i,s,r,j}^{(m)}
    =
    \frac{
        \left(
            W_q^{a}
            \mathbf{a}_{i,s,r}^{(m)}
        \right)^{\top}
        \left(
            W_k^{a}
            \mathbf{x}_{i,s,j}
        \right)
    }{
        \sqrt{d_a}
    },
    \qquad
    p_{i,s,r}^{(m)}(j)
    =
    \operatorname{softmax}_{j \in \mathcal{I}_m(i,s)}
    \left(
        \ell_{i,s,r,j}^{(m)}
    \right),
\label{eq:app-anchor-probe}
\end{equation}
where $W_q^{a}$ and $W_k^{a}$ are learnable probe projections and $d_a$ is the probe dimension. The distribution $p_{i,s,r}^{(m)}(j)$ represents where anchor $r$ looks within modality $m$.

Let $\mathcal{E}_{i,s}^{m}$ denote the structured evidence targets for modality $m$. Each evidence target $e \in \mathcal{E}_{i,s}^{m}$ is aligned to the serialized model input and converted into a normalized token-level distribution $y_e(j)$. The matching cost between anchor $r$ and evidence target $e$ is
\begin{equation}
    \operatorname{CE}_{i,s,r,e}
    =
    -
    \sum_{j \in \mathcal{I}_m(i,s)}
    y_e(j)
    \log
    p_{i,s,r}^{(m)}(j).
\label{eq:app-anchor-ce}
\end{equation}

Since evidence targets are not pre-assigned to specific anchors, DME uses a softmin assignment:
\begin{equation}
    \pi_{i,s,r|e}
    =
    \operatorname{softmax}_{r}
    \left(
        -
        \operatorname{CE}_{i,s,r,e}
        /
        \tau_{\mathrm{anc}}
    \right),
\label{eq:app-softmin-assignment}
\end{equation}
where $\tau_{\mathrm{anc}}$ is the assignment temperature. The assignment weights are detached when computing the evidence-level hit loss:
\begin{equation}
    \mathcal{L}_{i,s,e}^{\mathrm{hit}}
    =
    \sum_{r=1}^{R_m}
    \operatorname{sg}
    \left(
        \pi_{i,s,r|e}
    \right)
    \operatorname{CE}_{i,s,r,e},
\label{eq:app-evidence-hit}
\end{equation}
where $\operatorname{sg}(\cdot)$ denotes stop-gradient. This formulation allows different anchors to specialize to different evidence targets without imposing a hard assignment.

To prevent all evidence targets from being assigned to the same anchor, we define the average assignment mass
\begin{equation}
    u_{i,s,r}^{(m)}
    =
    \frac{1}{
        \left|
            \mathcal{E}_{i,s}^{m}
        \right|
    }
    \sum_{e \in \mathcal{E}_{i,s}^{m}}
    \pi_{i,s,r|e},
\label{eq:app-assignment-mass}
\end{equation}
and use the balance regularizer
\begin{equation}
    \mathcal{L}_{\mathrm{bal}}
    =
    \mathbb{E}_{
        \substack{
            i,\,
            s \in \mathcal{S}_{\mathrm{anc}},\,
            m:\\
            |\mathcal{E}_{i,s}^{m}|>0
        }
    }
    \left[
        R_m
        \sum_{r=1}^{R_m}
        \left(
            u_{i,s,r}^{(m)}
        \right)^2
        -
        1
    \right].
\label{eq:app-balance-loss}
\end{equation}
The complete anchor hit loss is
\begin{equation}
    \mathcal{L}_{\mathrm{hit}}
    =
    \mathbb{E}_{
        (i,s,m,e)
        \in
        \Omega_{\mathrm{hit}}
    }
    \left[
        \mathcal{L}_{i,s,e}^{\mathrm{hit}}
    \right]
    +
    \lambda_{\mathrm{bal}}
    \mathcal{L}_{\mathrm{bal}},
\label{eq:app-complete-hit-loss}
\end{equation}
where
\begin{equation}
    \Omega_{\mathrm{hit}}
    =
    \left\{
        (i,s,m,e):
        s \in \mathcal{S}_{\mathrm{anc}},
        \,
        e \in \mathcal{E}_{i,s}^{m}
    \right\}.
\end{equation}

\subsection{Summary-Conditioned Evidence Supervision}
\label{app:summary-conditioned-evidence}

The anchor distribution also defines an anchor-specific evidence representation:
\begin{equation}
    \mathbf{e}_{i,s,r}^{(m)}
    =
    \sum_{j \in \mathcal{I}_m(i,s)}
    p_{i,s,r}^{(m)}(j)
    \mathbf{x}_{i,s,j}.
\label{eq:app-anchor-evidence-pool}
\end{equation}
The side-level evidence pool is obtained by aggregating all available modality-specific anchor representations:
\begin{equation}
    \mathbf{e}_{i,s,\mathrm{pool}}
    =
    \operatorname{Mean}
    \left(
        \left\{
            \mathbf{e}_{i,s,r}^{(m)}
        \right\}_{m,r}
    \right).
\label{eq:app-side-evidence-pool}
\end{equation}

Let $\mathbf{s}_{i,s}$ denote the cached embedding of the teacher-generated local summary. DME uses
\begin{equation}
    \mathcal{L}_{\mathrm{sum}}
    =
    \mathbb{E}_{
        (i,s)
        \in
        \Omega_{\mathrm{sum}}
    }
    \left[
        1
        -
        \cos
        \left(
            W_{\mathrm{sum}}
            \mathbf{e}_{i,s,\mathrm{pool}},
            \mathbf{s}_{i,s}
        \right)
    \right],
\label{eq:app-summary-loss}
\end{equation}
where
\begin{equation}
    \Omega_{\mathrm{sum}}
    =
    \left\{
        (i,s):
        s \in \mathcal{S}_{\mathrm{anc}}
        \ \text{and}\
        \mathbf{s}_{i,s}
        \ \text{is available}
    \right\}.
\end{equation}
This objective encourages the evidence pool not only to cover teacher-annotated positions, but also to preserve their intended local semantics.

\subsection{Typed Latent Objectives}
\label{app:typed-latent-objectives}

For query $i$, let $\mathbf{h}_{i,k}^{\mathrm{traj}}$ denote the hidden state of the $k$-th typed latent token. We project it into the retrieval space:
\begin{equation}
    \mathbf{q}_{i,k}^{\mathrm{traj}}
    =
    \operatorname{norm}
    \left(
        P_{\mathrm{traj}}
        \mathbf{h}_{i,k}^{\mathrm{traj}}
    \right).
\label{eq:app-trajectory-projection}
\end{equation}

Each teacher-generated trajectory step has a type $\kappa_{i,k}$ and a state description whose cached embedding is denoted by $\mathbf{g}_{i,k}$. We define
\begin{align}
    \Omega_{\mathrm{sem}}
    &=
    \left\{
        (i,k):
        \kappa_{i,k}
        \in
        \mathcal{Y}_{\mathrm{sem}}
    \right\},
    \\
    \Omega_{\mathrm{align}}
    &=
    \left\{
        (i,k):
        \kappa_{i,k}
        \in
        \mathcal{Y}_{\mathrm{align}}
    \right\},
    \\
    \Omega_{\mathrm{reject}}
    &=
    \left\{
        (i,k):
        \kappa_{i,k}
        \in
        \mathcal{Y}_{\mathrm{reject}}
        \ \text{and}\
        d_i^-
        \ \text{is available}
    \right\}.
\end{align}
In the default setting, semantic supervision is applied to localization-oriented states, positive-alignment supervision is applied to \texttt{align\_pos} states, and rejection supervision is applied to \texttt{reject\_neg} states.

For semantic states, DME aligns the latent hidden state with the cached teacher-state embedding:
\begin{equation}
    \mathcal{L}_{\mathrm{sem}}
    =
    \mathbb{E}_{
        (i,k)
        \in
        \Omega_{\mathrm{sem}}
    }
    \left[
        1
        -
        \cos
        \left(
            P_{\mathrm{traj}}
            \mathbf{h}_{i,k}^{\mathrm{traj}},
            \mathbf{g}_{i,k}
        \right)
    \right].
\label{eq:app-semantic-state-loss}
\end{equation}

For positive-alignment states, the projected trajectory representation is trained to retrieve its corresponding positive document from the gathered positive-document bank:
\begin{equation}
    \mathcal{L}_{\mathrm{align}}
    =
    -
    \mathbb{E}_{
        (i,k)
        \in
        \Omega_{\mathrm{align}}
    }
    \log
    \frac{
        \exp
        \left(
            \mathbf{q}_{i,k}^{\mathrm{traj}\top}
            \operatorname{sg}
            \left(
                \mathbf{z}_{d_i^+}
            \right)
            /
        \tau_{\mathrm{traj}}
        \right)
    }{
        \sum_j
        \exp
        \left(
            \mathbf{q}_{i,k}^{\mathrm{traj}\top}
            \operatorname{sg}
            \left(
                \mathbf{z}_{d_j^+}
            \right)
            /
        \tau_{\mathrm{traj}}
        \right)
    },
\label{eq:app-positive-alignment-loss}
\end{equation}
where $\tau_{\mathrm{traj}}$ is the trajectory contrastive temperature.

For rejection states, DME applies a margin-ranking objective:
\begin{equation}
    \mathcal{L}_{\mathrm{reject}}
    =
    \mathbb{E}_{
        (i,k)
        \in
        \Omega_{\mathrm{reject}}
    }
    \left[
        \max
        \left(
            0,\,
            \mu
            +
            \cos
            \left(
                \mathbf{q}_{i,k}^{\mathrm{traj}},
                \operatorname{sg}
                \left(
                    \mathbf{z}_{d_i^-}
                \right)
            \right)
            -
            \cos
            \left(
                \mathbf{q}_{i,k}^{\mathrm{traj}},
                \operatorname{sg}
                \left(
                    \mathbf{z}_{d_i^+}
                \right)
            \right)
        \right)
    \right],
\label{eq:app-negative-rejection-loss}
\end{equation}
where $\mu$ is the rejection margin.

The complete typed latent objective is
\begin{equation}
    \mathcal{L}_{\mathrm{typed}}
    =
    \lambda_{\mathrm{sem}}
    \mathcal{L}_{\mathrm{sem}}
    +
    \lambda_{\mathrm{align}}
    \mathcal{L}_{\mathrm{align}}
    +
    \lambda_{\mathrm{reject}}
    \mathcal{L}_{\mathrm{reject}}.
\label{eq:app-typed-latent-loss}
\end{equation}

Combining the evidence and typed latent terms, the Stage-2A objective is
\begin{equation}
    \mathcal{L}_{\mathrm{S2A}}
    =
    \lambda_{\mathrm{hit}}
    \mathcal{L}_{\mathrm{hit}}
    +
    \lambda_{\mathrm{sum}}
    \mathcal{L}_{\mathrm{sum}}
    +
    \mathcal{L}_{\mathrm{typed}}.
\label{eq:app-stage2a-objective}
\end{equation}

\section{Visualization of Cross-Conditional Reconstruction}
\label{app:generation-vis}

We provide a qualitative view of what Stage 2-B injects into the
embedding in this section. Our goal is to verify that, after jointly optimizing the contrastive retrieval loss and the cross-conditional reconstruction loss, DME does not collapse the generative understanding of the MLLM backbone into a pure similarity geometry. Instead, the retrieval embedding remains \emph{language-decodable}: a single dense vector can be decoded back into text that is semantically faithful to its own input and, crucially, aligned with the counterpart it is matched against. This directly probes the central claim of Section~\ref{sec:stage2b-cross-conditional-self-decoding}, namely that the embedding is supervised to preserve fine-grained counterpart-side semantics.

\begin{figure}[p]
    \centering
    \includegraphics[width=1\linewidth]{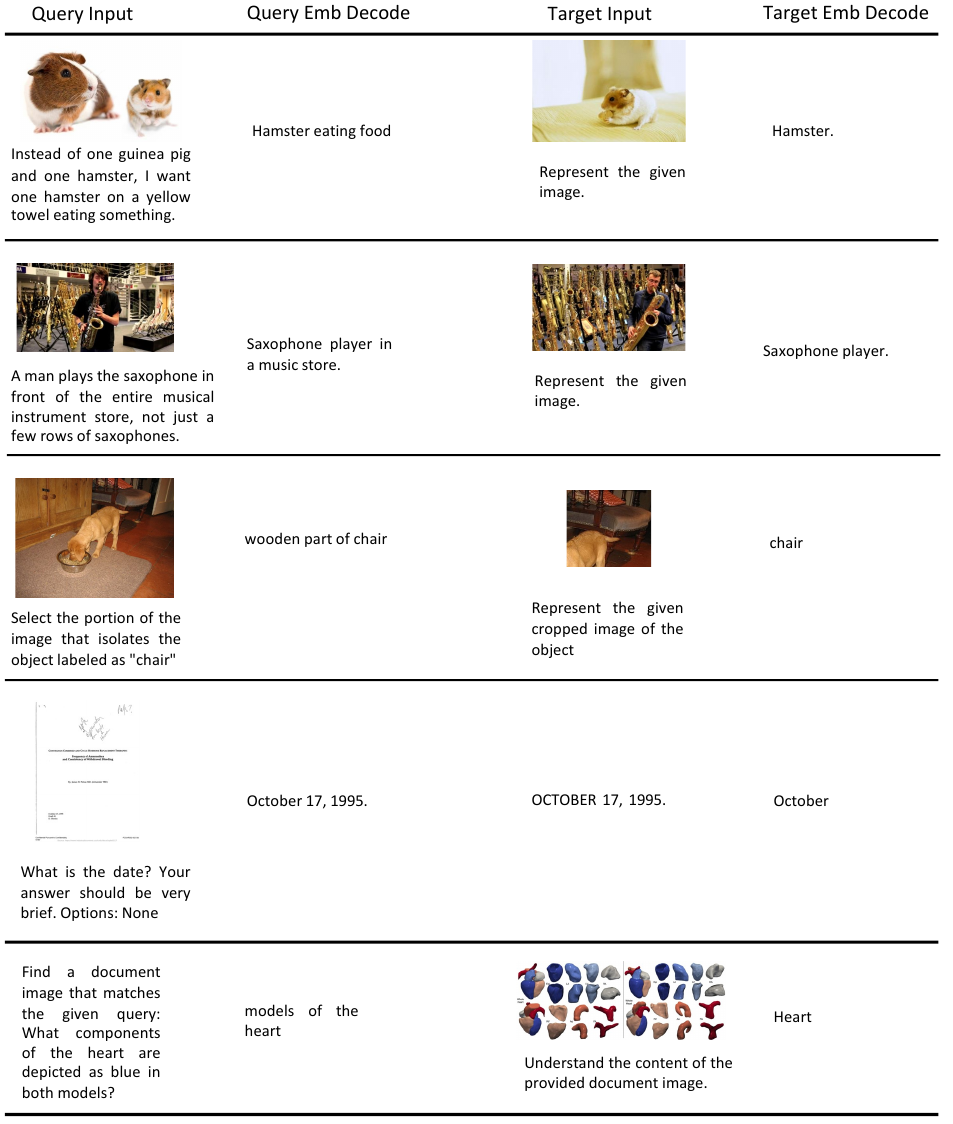}
      \caption{\textbf{Reconstruction visualization on image and visual-document inputs.}}
  \label{fig:generation_1}
\end{figure}

\begin{figure}[p]
    \centering
    \includegraphics[width=\linewidth]{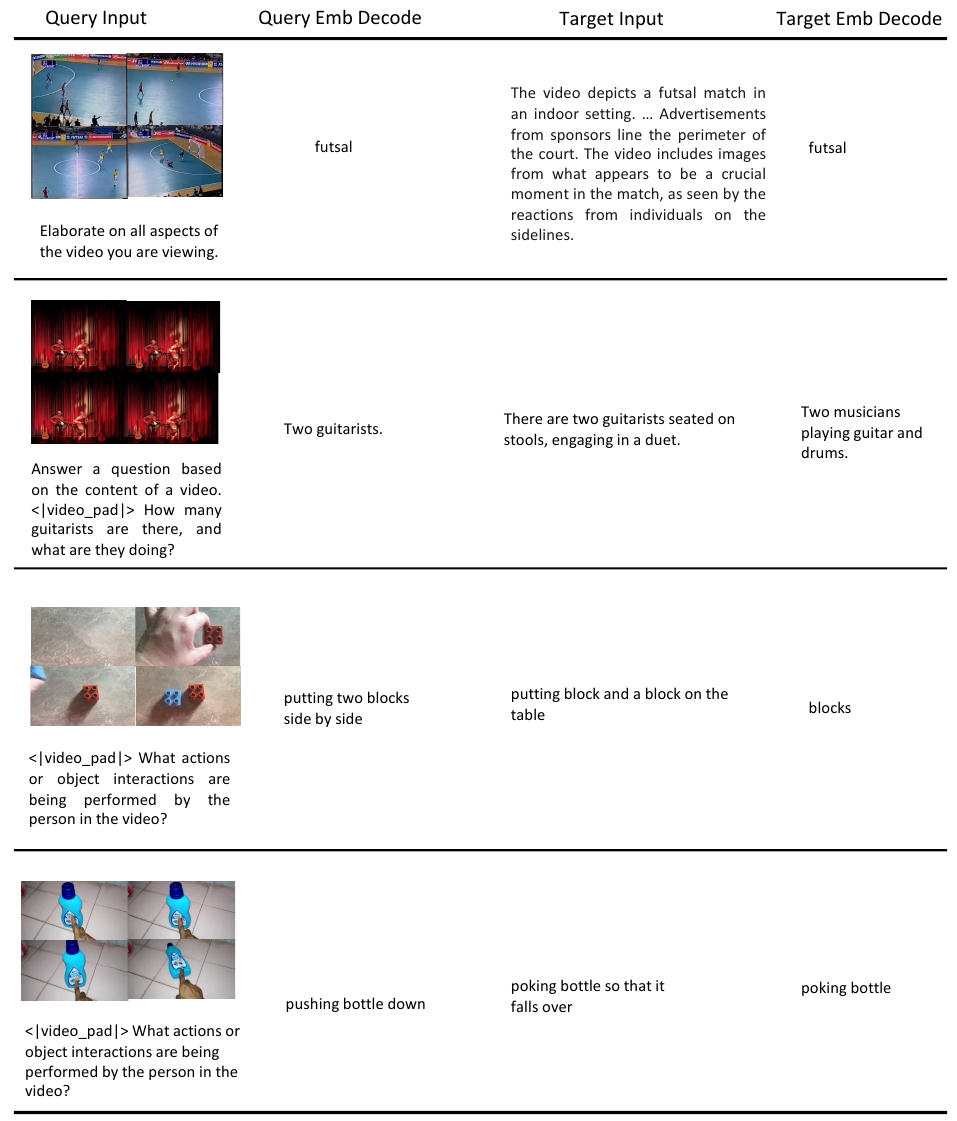}
  \caption{\textbf{Reconstruction visualization on video inputs.}}
    \label{fig:generation_2}
\end{figure}

\begin{figure}[p]
    \centering
    \includegraphics[width=\linewidth]{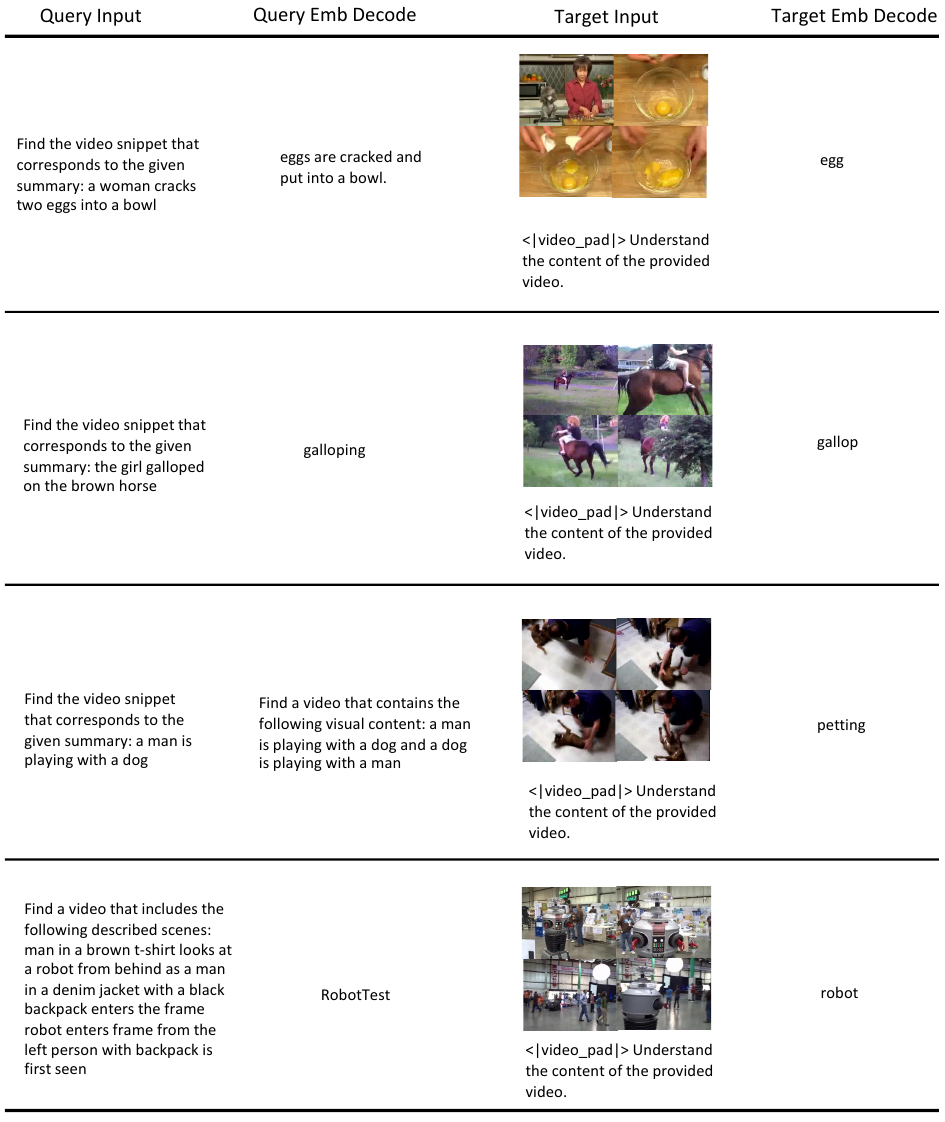}
    \caption{\textbf{Reconstruction visualization on video moment-retrieval inputs.}}
    \label{fig:generation_3}
\end{figure}

\paragraph{Reconstruction setup.}
For each query or document, we take the pre-normalization embedding $\tilde{\mathbf{z}}$ defined in Eq.~\eqref{eq:s2b-prefix} and use it as the \emph{only} prefix token fed back into the shared backbone $F_\theta$. No original text or visual tokens of the input are provided during the process: all decoded content is reconstructed purely from the single embedding. We use greedy decoding and keep a single output sequence per embedding. Under this protocol, the ``Query Emb Decode'' and ``Target Emb Decode'' columns in Figures~\ref{fig:generation_1}--\ref{fig:generation_3} are produced solely from the query-side and document-side embeddings, respectively.

\paragraph{Analysis.}
The visualization results are shown in Figures~\ref{fig:generation_1}, \ref{fig:generation_2}, and \ref{fig:generation_3}, covering image and visual-document inputs, video inputs, and video moment-retrieval inputs, respectively. We make the following observations.

First, both the query-side and the document-side embeddings decode into coherent and on-topic text. Since the cross-conditional reconstruction objective is applied symmetrically in the Q$\rightarrow$D and D$\rightarrow$Q directions, both encoding directions retain a decodable generative representation rather than only the query side. Moreover, many embeddings originate from purely visual or video inputs yet still decode into fluent text, indicating that the embedding stays grounded in a language-decodable semantic space and that the generative understanding of the backbone is preserved rather than collapsed by contrastive training.

Second, the decoded text is consistently much shorter than the raw input. For example, a long instruction-conditioned query is decoded into a compact phrase such as ``Hamster eating food'', and a full document image is decoded into a few salient words. This shows that the embedding behaves as an abstractive semantic bottleneck: it preserves the salient, retrieval-relevant gist of the input while discarding surface detail, rather than performing verbatim reconstruction.

Third, and most importantly, the embedding tends to surface the \emph{intersection} of the query and target semantics. Across examples, the query and target embeddings decode into overlapping core content---for instance, ``Hamster eating food'' versus ``Hamster'', or ``Saxophone player in a music store'' versus ``Saxophone player''---so that the shared concept dominates both decodings. This indicates that the model effectively learns the common, relevance-bearing features between a query and its target. Because retrieval relevance is itself defined over the shared semantics of a matched pair, an embedding that explicitly surfaces this intersection is well aligned with the relevance criterion, which in turn helps explain the retrieval gains brought by Stage 2-B.

Finally, we note that since decoding is greedy and lossy, fine-grained tokens are not always reconstructed exactly (\eg, ``October 17, 1995'' may be decoded only as ``October''). This is expected for a compact embedding and is consistent with our goal: the reconstruction objective is designed to enforce \emph{semantic} fidelity of counterpart-side content in the embedding, not lossless textual reconstruction.

\end{document}